\documentclass[journal]{IEEEtran}
\usepackage{graphics} 
\usepackage{epsfig} 
\usepackage{times} 
\usepackage{amsmath} 
\usepackage{amsthm}
\usepackage{booktabs}
\usepackage{amssymb}
\usepackage{tcolorbox}
\usepackage{multirow}
\usepackage{longtable}
\usepackage {threeparttable}
\usepackage{mathtools}
\usepackage[utf8]{inputenc}
\usepackage{tikz}
\usetikzlibrary{arrows.meta}
\usetikzlibrary{graphs}
\usepackage[colorlinks=false, urlcolor=blue, pdfborder={0 0 0}]{hyperref}
\usepackage{enumerate}
\usepackage{color}
\usepackage[linesnumbered,boxed,ruled,commentsnumbered]{algorithm2e}
\usepackage{subfigure}
\usepackage{comment}
\usepackage{cite}
\usepackage{flushend} \usepackage{float} 
\newfloat{tcolorboxenv}{h}{lob}

\definecolor{myred}{RGB}{182,20,50}
\definecolor{myblue}{RGB}{227, 245, 250}
\definecolor{myyellow}{RGB}{255,255,0}
\definecolor{mygreen}{RGB}{250, 250, 235}
\definecolor{myorange}{RGB}{255,128,0}
\definecolor{mygray}{RGB}{192,192,192}
\newtheorem{mydef}{Definition}
\newtheorem{mythm}{Theorem}
\newtheorem{myprob}{Problem}
\newtheorem{mylem}{Lemma}

\newtheorem{mypro}{Proposition}
\newtheorem{mycla}{Claim}
\newtheorem{myexm}{Example}
\newtheorem{remark}{Remark}
\usetikzlibrary{patterns}
\usetikzlibrary{shapes}

\def \M{\mathcal{M}}
\def \S{\mathcal{S}}
\def \A{\mathcal{A}}

\def \T{\mathcal{T}}

\title{Entropy Rate Maximization of Markov Decision Processes under Linear Temporal Logic Tasks}

\author{
Yu Chen, Shaoyuan Li and Xiang Yin
\thanks{This work was supported by the  National Natural Science Foundation of China (62173226, 62061136004, 61833012).}
	\thanks{Yu Chen, Shaoyuan Li and Xiang Yin are with Department of Automation and Key Laboratory of System Control and Information Processing, Shanghai Jiao Tong University, Shanghai 200240, China.
	{\tt\small \{yuchen26, syli, yinxiang\}@sjtu.edu.cn}. (Corresponding author: Xiang Yin)
}
}

\IncMargin{0.8em}
\begin{document}

\maketitle

\begin{abstract}
We investigate the problem of synthesizing optimal control policies for Markov decision processes (MDPs) with both qualitative and quantitative objectives. 
Specifically, our goal is to achieve a given linear temporal logic (LTL) task with probability one, while 
maximizing the \emph{entropy rate} of the system. 
The notion of entropy rate characterizes the long-run average (un)predictability of a stochastic process. 
Such an optimal policy is of our interest, in particular, from the security point of view, as it not only ensures the completion of tasks, but also maximizes the unpredictability of the system.
However, existing works only focus on maximizing the total entropy  which may diverge to infinity for infinite horizon. 
In this paper, we provide a complete solution to  the entropy rate maximization problem under LTL constraints. 
Specifically, we first present an algorithm for synthesizing entropy rate maximizing policies for communicating MDPs.  Then based on a new state classification method, we show the entropy rate maximization problem under LTL task can be effectively solved in polynomial-time.  We illustrate the proposed algorithm based on two case studies of robot task planning scenario. 
\end{abstract}

\begin{IEEEkeywords}
Markov Decision Process, Entropy Rate, Linear Temporal Logic, Security.
\end{IEEEkeywords}

\IEEEpeerreviewmaketitle

\section{Introduction}

\subsection{Motivations}
\IEEEPARstart{T}{ask}  planning and decision-making are central problems in autonomous systems. 
For example, a field mobile robot needs to navigate in intricate and uncertain environments while achieving complex tasks associated with its spatiotemporal behavior.
Markov Decision Processes (MDPs) provide a foundational mathematical framework for decision-making under uncertainty in this context. By abstracting system dynamics as well as uncertainties as transition probabilities, MDPs offer a suitable tool for modeling and optimizing the interactions between the decision-maker and the environment, enabling adaptability and efficient response to ever-changing conditions.

In the context of MDPs, numerous works have focused on synthesizing optimal control policies that maximize or minimize various quantitative performance metrics, such as total reward, long-run average reward (or mean payoff), and discounted total reward \cite{puterman}.
More recently, motivated by the growing needs in decision-making for complex requirements in autonomous systems, optimal control for MDPs with high-level tasks described by temporal logic formulae has also drawn considerable attention in the literature; see, e.g.,  
 the recent survey papers \cite{kress2018synthesis,belta2019formal,yin2024formal}. Among many formal specification languages, linear temporal logic (LTL)~\cite{baier2008principles} offers a rich and user-friendly way for designers to describe desired high-level specifications, such as ``\emph{visiting region $A$ infinitely often while never reaching region $B$}".  
For example, when the structural information of MDP models is known, algorithms have been developed to synthesize optimal policies achieving LTL tasks with probabilistic guarantees \cite{ding2014optimal, guo2018probabilistic, niu2019optimal}. When transition probabilities in MDPs are unknown a priori, reinforcement learning techniques have also been used to learn optimal policies for LTL tasks \cite{hahn2019omega, cai2021modular, voloshin2022policy}.

Despite the randomized nature of MDPs, the control policy applied can also be randomized, i.e., at each time instant, the decision-maker will choose a specific action according to some probability distribution. To accomplish an LTL task, it is known that considering deterministic policies is already sufficient~\cite{baier2008principles}. Yet, using randomized policies still have many advantages beyond satisfaction consideration. One important reason is the security consideration in adversarial environments.
For example, when the transition probability of each action in an MDP is one, applying a deterministic policy results in a purely deterministic trajectory. In this case, a malicious attacker could easily hack the information of the system or even physically destroy the agent by predicting its trajectory. Therefore, synthesizing randomized policies that not only achieve the desired task but also make the system's behavior as \emph{unpredictable} as possible is of great importance and has drawn considerable attention from researchers recently \cite{hibbard2019unpredictable,li2020detection,duan2021markov,zheng2022privacy}.  
For example, several different measures for unpredictability have been used such as  pre-opacity \cite{yang2023secure,chen2023you}, Bayes risk~\cite{CHATZIKOKOLAKIS2008378} and entropy~\cite{paruchuri2006security,NEKOUEI2019412}.
  
\subsection{Our Results}
In this work, we investigate the problem of unpredictable control policy synthesis under LTL task constraints for stochastic systems modeled as finite MDPs. 
Among many different notions of unpredictability of the system's behavior, \emph{entropy} is a widely used and very fundamental information-theoretic measure for quantifying how uncertain a random variable or a stochastic process is~\cite{thomas2006elements}. 
Here, we adopt the notion of \emph{entropy rate} as a measure of the unpredictability for the infinite horizon behavior of the system. Specifically, the entropy rate characterizes the long-run average total entropy of a stochastic process; hence, a higher entropy rate implies greater unpredictability of the stochastic process.
Our objective is to synthesize an optimal policy in the sense that 
 (i) the given LTL task is satisfied with probability one (w.p.1); and 
(ii) the entropy rate of the induced stochastic process  is maximized. 

Our approach for solving the entropy rate maximization problem under LTL constraints consists of two stages. 
First, we restrict our attention to the special case of communicating MDPs. For this case, we reveal a new structural property that shows the optimal policy is stationary and its resulting Markov chain is irreducible. Based on this structural property, a convex program is formulated to find the optimal policy.
Then, we tackle the general case where the MDPs may not be communicating. To this end, we introduce a novel state classification technique that characterizes system states into different levels. 
The overall synthesis algorithm iteratively solves convex programs for communicating MDPs within each level as well as
a new linear program that determines how high-level states transition to lower levels.
We prove that our synthesis algorithm is both sound and complete. Furthermore, its computational complexity is polynomial with respect to the size of the MDP. 
Moreover, we have implemented our algorithm, and two case studies on robot task planning are provided to demonstrate the effectiveness of our approach.

\subsection{Related Works}
In the past years, formal control synthesis for MDPs under LTL tasks has drawn much attention in the literature. For example, in \cite{chatterjee2015measuring, guo2018probabilistic}, the authors investigate how to synthesize optimal policies that achieve LTL tasks while maximizing the long-run average reward (or mean payoff). In \cite{ding2014optimal}, the authors further consider the optimal control problem for a new metric called average reward per cycle. 
In \cite{wolff2012robust}, the authors consider the robust control problem for uncertain MDPs. 
However, these metrics only characterize  the optimality of the policy rather than the unpredictability. Therefore, the optimization problems considered therein are very different from the entropy rate optimization problem.

In the context of unpredictable policy synthesis, \cite{BIONDI2014384} first shows how to maximize the total entropy of MDPs. This result is further extended to the partial observation setting in \cite{savas2022entropy}. However, in general, the total entropy of a stochastic process diverges unless the process becomes deterministic eventually. Therefore, for MDPs with infinite horizon, recent works have considered the maximization of \emph{entropy rate} \cite{burda2009localization, LIPIcs:2014:4872, george2018markov}, which is the long-run average of the total entropy. For example, \cite{LIPIcs:2014:4872} computes the value of the maximum entropy rate for MDPs without any task constraint. In \cite{1056939}, the authors consider the maximization of entropy rate under moments constraints. In \cite{george2018markov}, the authors further consider  stationary distributions as constraints in addition to entropy rate maximization. However, none of the above-mentioned works consider the temporal logic requirements when maximizing the unpredictability of the MDP. Moreover, \cite{1056939, burda2009localization, george2018markov} only consider communicating MDPs, which is a special case of the general MDPs we consider in this work.

Our work is mostly related to \cite{savas2019entropy}, which solves the problem of maximizing the total entropy under LTL constraints. However, as we mentioned, the total entropy generally goes to infinity for systems operating over infinite horizon. Therefore, the approach in \cite{savas2019entropy} essentially yields a solution in which the steady state of the system is deterministic, which is somewhat restrictive. That is, only unpredictability in the transient part is taken into account, and the steady part is, in fact, completely predictable. In fact, since LTL formulae are evaluated over infinite traces, it seems more natural to consider entropy rate rather than total entropy as the metric for unpredictability when dealing with LTL tasks.

Finally, we note that entropy maximization has also been widely adopted in the context of reinforcement learning as a regularization term in the objective function; see, e.g., \cite{lazic2020maximum,neu2017unified,cen2022fast}. Specifically, entropy maximization can be used to promote exploration and accelerate the convergence of the learning agent. However, our purpose for entropy rate maximization here is to enhance the unpredictability of the agent's behavior. Moreover, these works consider model-free learning problems, while  here we consider a model-based optimal control problem.

\subsection{Organization}
The remaining parts of the paper are organized as follows. 
We first introduce some necessary preliminaries in Section~\ref{sec:2} and then formulate the problem of entropy rate maximization under LTL constraints in Section~\ref{sec:3}. In Section~\ref{sec:4}, we solve the problem by focusing on a special class of MDPs called communicating MDPs. Next, in Section~\ref{sec:5}, we introduce the technique of state-level classification  and show how to leverage this technique to solve the problem for the general case in Section~\ref{sec:6}. Two case studies on robot task planning are provided in Section~\ref{sec:7} to illustrate our results. Finally, we conclude the paper in Section~\ref{sec:8}.

The preliminary version of some results in this paper is presented in~\cite{chen2023entropy}. 
Compared with~\cite{chen2023entropy}, the present work has the following differences.  
First and foremost, this paper considers the general LTL tasks while  \cite{chen2023entropy} only considers the so called surveillance tasks. 
Furthermore, this paper contains detailed proof compared to the sketchy analysis in \cite{chen2023entropy}. Finally, additional experimental results are provided to illustrate the effectiveness of our approach.  

\section{Preliminary}\label{sec:2}

\subsection{Markov Decision Processes}
A (finite and labeled) Markov decision process (MDP)  is a $6$-tuple
\begin{equation}
 \mathcal{M} = (S,s_0,A,P,\mathcal{AP},\ell), 
 \end{equation}
 where 
 $S=\{1,\dots,n\}$ is a finite set of states, 
 $s_0 \in S$ is the initial state,
 $A$ is a finite set of actions, 
 $P : S \times A \times S \rightarrow [0,1]$ is a transition function such that: 
 for any  $s \in S, a \in A$, we have 
 $\sum_{s' \in S} P(s'\mid s,a)\in \{0,1\}$,
 $\mathcal{AP}$ is a set of atomic propositions, and
 $\ell:S \to 2^{\mathcal{AP}}$ is a labeling function that assigns each state a set of atomic propositions.
 For simplicity, we also write $ P(s'\mid s,a)$ as $P_{s,a,s'}$.  
 We denote by $\textsf{succ}(s,a)=\{ s' \mid  P(s'\mid s,a) > 0 \}$ the set of successor states of state $s\in S$ under action $a\in A$. 
 For each state $s\in S$, we denote by $A(s)=\{a\in A  \mid \textsf{succ}(s,a)\neq \emptyset\}$ the set of available actions at $s$. 
 We assume that, for each state, there  exists at least one available action, i.e., $\forall s\in S: A(s)\neq \emptyset$. 
 An MDP also induces an underlying directed graph (digraph), 
 where each vertex is a state and an edge of form $\langle s,s'\rangle$ is defined whenever $P(s'\mid s,a)>0$ for some $a\in A$.  
  We denote by  $\pi_0$ the \emph{initial distribution} such that $\pi_0(s)=1$ if $s=s_0$ is the initial state and $\pi_0(s)=0$ otherwise.

 A Markov chain (MC) $\mathcal{C}$ is an MDP such that $|A(s)|=1$ for all $s \in S$. 
 We denote by  $\mathbb{P}$ the \emph{transition matrix} of an MC, i.e.,   $\mathbb{P}_{s,s'}=P(s'\mid s,a)$, where $a\in A(s)$ is the unique action at state $s\in S$.  
 Therefore, we can omit action set and write an MC as $\mathcal{C}=(S,s_0,\mathbb{P})$.
 The \emph{limit transition matrix} of an MC is defined by
 $\mathbb{P}^\star=\lim_{n \rightarrow \infty} \frac{1}{n} \sum_{k=0}^{n} \mathbb{P}^{k}$. Note that this limit matrix always exists for any finite MC \cite{puterman}.

A \emph{policy} for an MDP $\mathcal{M}$ is a sequence 
$\mu = (\mu_{0}, \mu_{1},... )$,  where each $\mu_{k} : S \times A \rightarrow [0,1]$ is a function such that 
$\forall s\in S: \sum_{a \in A(s)} \mu_{k}(s,a)=1$. 
A policy is said to be \emph{stationary} if $\mu_{i}=\mu_{j}$ for all $i,j$ 
and we write a stationary policy by $\mu=(\mu,\mu,\dots)$ for simplicity. 
Given an MDP $\mathcal{M}$, the sets of all policies and all stationary policies are denoted by $\Pi_\mathcal{M}$ and $\Pi^{S}_\mathcal{M}$, respectively. 
For policy $\mu \in \Pi_{\M}$, at time $k$, $\mu$ induces a  transition matrix $\mathbb{P}^{\mu_k}$ such that $\mathbb{P}^{\mu_k}_{i,j}=\sum_{a\in A(i)}\mu_k(i,a)P_{i,a,j}$. 
 We denote by $\textsf{reach}(s)$ the set of all states reachable from state $s\in S$ in MDP, i.e., 
\[
\textsf{reach}(s)=\{ s' \in S \mid  \exists \mu \in \Pi^S_{\mathcal{M}}, \exists n\in \mathbb{N} \text{ s.t. } (\mathbb{P}^{\mu})^{n}_{s,s'} > 0  \}.
\]

Here $(\mathbb{P}^\mu)^n_{s,s'}$ is the transition probability from $s$ to $s'$ in transition matrix $(\mathbb{P}^\mu)^n$,  which is the $n$-th power of transition matrix $\mathbb{P}^\mu$.
 
 Let $\mathcal{M}= (S,s_0,A,P,\mathcal{AP},\ell)$ be an MDP and  $\mu \in \Pi_{\M}$ be a policy. 
 An infinite sequence $\rho=s_{0}s_{1}\cdots$ of states is said to be a \emph{path} in $\mathcal{M}$ under $\mu$
 if (i) $s_0$ is the initial state of the MDP, and (ii) $\forall k\geq 0:\sum_{a\in A(s_{k})}   \mu_{k}(s_k,a) P(s_{k+1}\mid s_{k},a) > 0 $. 
 We denote by $\textsf{Path}^{\mu}(\mathcal{M})\subseteq S^\omega$ the set of all paths in $\mathcal{M}$ under $\mu$, where $ S^\omega$ denotes the set of all infinite sequences of symbols over set $S$. 
 We use the standard probability measure 
 $\textsf{Pr}^{\mu}_{\mathcal{M}}: 2^{S^\omega}\to[0,1]$ for the sample space over  infinite paths, which satisfies: 
 for any finite sequence $s_0\cdots s_n$, we have
   \[
   \textsf{Pr}^{\mu}_{\mathcal{M}}(\textsf{Cly}(s_{0}\dots s_{n}) ) 
   \!=\!\pi_{0}(s_0) \prod_{k=0}^{n-1} \sum_{a \in A(s_{k})}
   \!\! \mu_{k}(s_{k},a) P_{s_k,a,s_{k+1}},
   \]  
   where $\textsf{Cly}(s_{0}\dots s_{n}) \subseteq \textsf{Path}^{\mu}(\mathcal{M})$ is  the \emph{cylinder 
} set, which is  the set of all paths having prefix $s_{0}\dots s_{n}$. 
   The reader is referred to   \cite{baier2008principles} for details on this probability measure on infinite paths.

Given an MDP, let 
$(\mathcal{S},\mathcal{A})$ be a state-action pair, where 
$\mathcal{S} \subseteq S$ is a non-empty set of states and 
$\mathcal{A} :\mathcal{S} \rightarrow 2^{A}\setminus \emptyset$ is a function such that 
 (i) $\forall s\in \mathcal{S}: \mathcal{A}(s) \subseteq A(s)$; and 
(ii) $\forall s\in \mathcal{S},a\in \mathcal{A}(s): \textsf{succ}(s,a)\subseteq \mathcal{S}$. 
Essentially, state-action pair $(\mathcal{S},\mathcal{A})$ induces a new MDP called the    \emph{sub-MDP} of $\mathcal{M}$, denoted by $\mathcal{M}(\mathcal{S},\mathcal{A})$ (or $(\mathcal{S},\mathcal{A})$ directly), by restricting the state space to $\mathcal{S}$ and available actions to $\mathcal{A}(s)$ for each state $s\in \mathcal{S}$.

\begin{mydef}[\bf Maximal End Components]\upshape
Let  $(\mathcal{S},\mathcal{A})$ be a sub-MDP of $\mathcal{M}= (S,s_0,A,P,\mathcal{AP},\ell)$. 
We say  $(\mathcal{S},\mathcal{A})$ is an \emph{end component} if its underlying digraph is strongly connected.  
We say $(\mathcal{S},\mathcal{A})$ is a  \emph{maximal end component} (MEC) if it is an end component and there is no other end component $(\mathcal{S}',\mathcal{A}')$  such that 
(i) $\mathcal{S}\subseteq \mathcal{S}'$; and (ii) $\forall s \in \S, \A(s) \subseteq \A'(s)$.
We denote by $\texttt{MEC}(\mathcal{M})$ the set of all MECs in $\mathcal{M}$.
\end{mydef}

Intuitively, if $(\mathcal{S},\mathcal{A})$ is an MEC, then we can find a policy such that, once a state in $\S$ is reached, we will stay in the MEC forever and all states in it will be visited infinitely w.p.1 thereafter. 
  
  \subsection{Entropy Rate of Stochastic Processes}
  Let $X$ be a discrete random variable with support $\mathcal{X}$ and $p(x):= \textsf{Pr}(X=x),x\in \mathcal{X}$ be its  probability mass function. The entropy of random variable $X$ is defined as:
  \begin{equation}
      H(X) := -\sum_{x \in \mathcal{X}} p(x)\log  p(x).  \label{(4)} 
  \end{equation}
 All logarithms in this work are with base 2 and we define $0\log (0) = 0$. 
  For two random variables $X_0$ and $X_1$ with joint probability mass function $p(x_0,x_1)$, 
  the \emph{joint entropy} of $X_0$ and $X_1$ is defined by 
  \begin{equation}
    H(X_{0},X_{1}) := -\sum_{x_{0} \in \mathcal{X}}\sum_{x_{1} \in \mathcal{X}} p(x_{0},x_{1})\log  p(x_{0},x_{1}).\label{(5)}
  \end{equation}
The joint entropy can also be directly extended to a discrete time stochastic process $\{X_k\}$. Intuitively, it provides a measure for how \emph{unpredictable} the process is. However, joint entropy $H(X_{0},X_{1},\dots,X_n)$ usually diverges when $n$ goes to infinity. Therefore, for infinite processes, one usually uses the \emph{entropy rate} instead of the joint entropy. 
\begin{mydef}[\bf Entropy Rate~\cite{thomas2006elements}]\upshape 
The entropy rate of a discrete-time stochastic process $\{X_k\}$ is defined as 
\begin{equation}
     \nabla H(\{X_k\}) := \lim_{k \rightarrow \infty} \frac{1}{k} H(X_{0},\dots,X_{k}) \label{(7)}
  \end{equation} 
  when the limit exists.
\end{mydef}  
Given an MC $\mathcal{C}=(S,s_0,\mathbb{P})$, it  also induces a discrete-time stochastic process $\{ X_{k}: k \in \mathbb{N} \}$, where $X_{k}$ is a random variable over state space $S$. 
 We denote by $\nabla H(\mathcal{C})$ the entropy rate of MC $\mathcal{C}$, which is the entropy rate of its induced process.   
 It is known that  this entropy rate can be computed by  \cite{LIPIcs:2014:4872}: 
   \begin{equation}
     \nabla H(\mathcal{C}) = \sum_{s\in S} \pi(s)L(s) , \label{(8)}
  \end{equation}
  where 
 $\pi = \pi_0 \mathbb{P}^{\star}$  is the limit distribution, and $L(s)$ is the so called \emph{local entropy} defined by 
  \begin{equation}
      L(s) = \sum_{s' \in S}-\mathbb{P}_{s,s'}\log  \mathbb{P}_{s,s'}. \label{def:localentropy}
  \end{equation}
 For MDP $\M$, we only consider polices under which \eqref{(7)} is well-defined and directly denote $\Pi_\M$ as such policies set. We define $\nabla H(\mathcal{M}):= \sup_{\mu \in \Pi_\mathcal{M}} \nabla H(\M^\mu)$ as its entropy rate. 

\subsection{Linear Temporal Logic}
We employ Linear Temporal Logic (LTL) to express formal tasks. 
Let $\mathcal{AP}$ be the set of atomic propositions. 
An LTL formula is constructed based on atomic propositions, Boolean operators and   temporal operators. Specifically, the syntax of LTL formulae is defined recursively as follows: 
\[
\varphi 
::= 
true  
\mid a 
\mid \varphi_1\wedge \varphi_2
\mid \neg \varphi
\mid \bigcirc \varphi
\mid \varphi_1 U \varphi_2,
\] 
where $a \in \mathcal{AP}$ is an atomic proposition; 
$\neg$ and $\wedge$ are Boolean operators ``negation" and ``conjunction", respectively; 
$\bigcirc$ and $U$ are temporal operators  ``next'' and ``until'', respectively. 
Note that one can further induce  temporal operators 
such as  ``eventually''  $\lozenge \varphi := true U \varphi$ 
and 
``always''  $\square\varphi := \neg \lozenge \neg \varphi$. 

An LTL formula $\varphi$ is interpreted over infinite words on $2^{\mathcal{AP}}$. 
Readers can find detailed information about the semantics of LTL formulae in \cite{baier2008principles}.
For any infinite word $\sigma \in (2^{\mathcal{AP}})^{\omega}$, 
we denote by $\sigma \models \varphi$ if it satisfies LTL formula $\varphi$. 
We denote by $\mathcal{L}_{\varphi}=\{ \sigma \in (2^{\mathcal{AP}})^{\omega}\mid 
\sigma \models \varphi \}$ the set of all infinite words satisfying  $\varphi$. 

\begin{mydef}[\bf Deterministic Rabin Automata]\upshape
A \emph{deterministic Rabin automata} (DRA) is a tuple $R=(Q,\Sigma,\delta,q_0,Acc)$, where $Q$ is a finite set of states, $\Sigma$ is a finite set of alphabet, $\delta: Q \times \Sigma \to Q$ is the transition function, $q_0 \in Q$ is the initial state, and $Acc=\{(B_1,G_1),\dots,(B_n,G_n) \}$ is a finite set of Rabin pairs such that $B_i,G_i \subseteq Q$ for all $i=1,2,\dots,n$.
\end{mydef}

Given an infinite word $\sigma=\sigma_1\sigma_2\cdots \in \Sigma^{\omega}$,  
its induced infinite \emph{run} in DRA $R$ is the sequence of states $\rho=q_0 q_1\cdots \in Q^{\omega}$ such that $q_i=\delta(q_{i-1},\sigma_i)$ for all $i \geq 1$. An infinite run $\rho \in Q^{\omega}$ is said to be accepted if  there exists a Rabin pair $(B_i,G_i) \in Acc$ such that $\textsf{inf}(\rho)\cap G_i\neq \emptyset$ and $\textsf{inf}(\rho) \cap B_i = \emptyset$, where $\textsf{inf}(\rho)$ is set of states that occur infinitely many times in $\rho$. 
An infinite word $\sigma$ is said to be \emph{accepted} if its induced  infinite run is accepted. 
We denote by $\mathcal{L}(R) \subseteq \Sigma^{\omega}$ the set of all accepted words of DRA $R$. 
Given an arbitrary LTL formula $\varphi$ over $\mathcal{AP}$, 
it is well-known that \cite{baier2008principles}, there exists a DRA with $\Sigma=2^{\mathcal{AP}}$ that accepts all infinite words satisfying $\varphi$, i.e., $\mathcal{L}_{\varphi}=\mathcal{L}(R)$.

Given an MDP $\M$ under policy $\mu$, a path $\rho=s_0s_1\cdots \in \textsf{Path}^{\mu}(\M)$   generates a word $\ell(\rho)=\ell(s_0)\ell(s_1)\cdots  \in (2^\mathcal{AP})^\omega$. 
Given an LTL formula $\varphi$, 
 we define
\[
\textsf{Pr}^{\mu}_{\M}(s_0 \models \varphi) :=\textsf{Pr}^{\mu}_{\M}(\{\rho \in \textsf{Path}^{\mu}(\M) \mid \ell(\rho) \models \varphi \})
\]
as the probability of satisfying LTL formula $\varphi$ for MDP $\M$ starting from $s_0$ under policy $\mu \in \Pi_{\M}$. 
We denote by $  \Pi^{\varphi}_{\M}$ as the set of policies under which the LTL task can be satisfied with probability 1, i.e., 
  \[
  \Pi^{\varphi}_{\M}=\{ \mu \in \Pi_{\M} \mid   \textsf{Pr}^{\mu}_{\M}(s_0 \models \varphi)=1\}.
  \]

\subsection{Product MDPs}
To integrate the task information into the MDP model, it is necessary to construct the product system between the DRA representing the LTL task and the original MDP.
 
\begin{mydef}[\bf Product MDPs]\upshape
Let $\M=(S,s_0,A,P,\mathcal{AP},$ $\ell)$ be an MDP and $R=(Q,2^{\mathcal{AP}},\delta,q_0,Acc)$ be the DRA accepting LTL formula $\varphi$. 
The \emph{product MDP} $\M_{\otimes}=(S_{\otimes},s_{0,{\otimes}},A,P_{\otimes},\mathcal{AP},\ell_{\otimes},Acc_{\otimes})$ is a 7-tuples,
where $S_{\otimes} = S \times Q$ is the product space, 
$s_{0,{\otimes}}=(s_0,q)$ is the initial state such that $q=\delta(q_0,\ell(s_0))$,
 $P_{\otimes}:S_{\otimes} \times A \times S_{\otimes} \to [0,1]$ is transition function defined by
  \begin{align}\label{eq:producttran}
	P_{\otimes}((s,q),a,(s',q')) \!= \!
		\left\{\!\!
		\begin{array}{cl}
			P_{s,a,s'} & \text{if }    q'=\delta(q,\ell(s'))  \\
			0                & \text{otherwise}  
		\end{array}
		\right.\!\!\!,   
\end{align} 
$\ell_{\otimes}$ is the labeling function such that
$\ell_{\otimes}((s,q))=\ell(s)$ and
$Acc_{\otimes}=\{ (B_1^{\otimes},G_1^{\otimes}),\dots,(B_n^{\otimes},G_n^{\otimes}) \}$ such that $B_i^{\otimes}=S \times B_i$ and $G_i^{\otimes}=S \times G_i$ for all $i=1,\dots n$.
\end{mydef}

Note that, since $R$ is deterministic, there exists a one-to-one correspondence between paths in $\M$ and $\M_{\otimes}$ \cite{baier2008principles}. 
Specifically, let  $s=s_0\dots s_n$ be a path executed in $\M$. 
Then there exists a unique path  $s'=s'_0\dots s'_n$ in $\M_{\otimes}$, where $s'_i=(s_i,q_i)$,  such that 
$s'_0=s_{0,{\otimes}}$ and  $q_i=\delta(q_{i-1},\ell(s_i)), i\geq 1$. 
Since the action spaces of $\M$ and $\M_{\otimes}$ are same, 
there also exists  a one-to-one correspondence between policies in $\M$ and $\M_{\otimes}$ 
\cite{guo2018probabilistic}.  

For the sake of simplicity,  hereafter in this paper,  we will omit the subscript and directly denote by $\M=(S,s_0,A,P,\mathcal{AP},\ell,Acc)$ the product MDP. The control synthesis problem is solved based on the product MDP. 
Specifically, for any path in (product) MDP $\rho=s_0s_1\dots$, it satisfies the LTL formula if and only if there exists an accepting pair $(B_{k},G_{k})\in Acc$ such that $\textsf{inf}(\rho)\cap G_{k} \neq \emptyset$ and $\textsf{inf}(\rho)\cap B_{k} = \emptyset$.  This accepting condition can be captured by the notion of accepting maximal end component. 

\begin{mydef}[\bf Accepting Maximal End Components]\upshape
Given a (product) MDP $\M=(S,s_0,A,P,\mathcal{AP},\ell,Acc)$, an \emph{accepting end component} (AEC) of product MDP is an end component $(\S,\A)$ such that for some accepting pair $(B_{k},G_{k})\in Acc$, 
we have $\S \cap B_{k} = \emptyset$ and $\S \cap G_{k} \neq \emptyset$. Moreover $(\S,\A)$ is said to be an \emph{accepting maximal end component} (AMEC) 
if there exists no other AEC $(\S',\A')$ such that 
(i) $\mathcal{S}\subseteq \mathcal{S}'$; and (ii) $\forall s\in \S, \A(s) \subseteq \A'(s)$. We denote by $\texttt{AEC}(\M)$ and $\texttt{AMEC}(\M)$ the set of AECs and AMECs of product MDP $\M$, respectively. 
\end{mydef}

Intuitively, given policy $\mu \in \Pi_{\M}$, the probability of satisfying a given LTL formula is equal to the probability of reaching AMEC and staying there forever. 
Note that both the set of all MECs and the set of all AMECs can be found effectively via graph search over the product space; see, e.g., \cite{baier2008principles,guo2018probabilistic}. 
Although states in different MECs are always disjoint,  this is not true for different AMECs. 
This is because different  AMECs may be accepted by different accepting pairs and their union may violate both acceptance conditions.

\section{Problem Formulation} \label{sec:3}
Now, we are ready to  formulate the problem that we solve in this paper.  
Our objective is to synthesize a control policy such that 
\begin{enumerate}
    \item 
    The given LTL task is satisfied with probability one; and 
    \item 
    The behavior of the agent needs to be as unpredictable as possible in the sense that the entropy rate of its induced stochastic process  is maximized. 
\end{enumerate}
This problem is formally stated as follow.
 
\begin{myprob}[\bf Entropy Rate Maximization for Linear Temporal Logic Tasks]\label{problem:maxicons} 
 Given MDP $\mathcal{M}$ and LTL formula $\varphi$, 
 which is equivalent to given the product MDP, 
 find an optimal policy $\mu^\star \in   \Pi^{\varphi}_{\M}$ such that
  \[
  \nabla H(\mathcal{M}^{\mu^\star})=\nabla H_{\varphi}(\mathcal{M}),
  \]
  where $\nabla H_{\varphi}(\mathcal{M})=\sup_{\mu \in   \Pi^{\varphi}_{\M}} \nabla H(\M^{\mu})$. 
\end{myprob}
Without loss of generality, we assume that, starting from each state in the product MDP, 
there exists a policy such that the LTL task can be satisfied w.p.1.  
Otherwise, we can use  Algorithm 45 in~\cite{baier2008principles} to eliminate such undesired states in polynomial time. 
The following result reveals a key structural property of Problem~\ref{problem:maxicons}, which shows that focusing only on stationary policies is sufficient.  
  \begin{mypro} \label{prop:suffisenouth} 
Let $\M=(S,s_0,A,P,\mathcal{AP},\ell,Acc)$ be a product MDP. Then if $\Pi^{\varphi}_{\M} \neq \emptyset$,  there exists $\mu^\star \in   \Pi^{\varphi}_{\M}\cap \Pi^{S}_{\M}$ such that
  \[
  \nabla H_\varphi(\M)=\sup_{\mu \in   \Pi^{\varphi}_{\M}\cap \Pi^{S}_{\M}} \nabla H(\M^{\mu})=\nabla H(\M^{\mu^\star}).
  \]
\end{mypro}
\begin{proof}
Due to space constraint, all proofs in the paper are omitted. They are available in the supplementary material \cite{chen2022entropy}.    
\end{proof}
Essentially, Proposition \ref{prop:suffisenouth} ensures that the Problem~\ref{problem:maxicons} has no solution iff  $\Pi_{\mathcal{M}}^\varphi=\emptyset$. 
To provide a complete solution, we only need to check whether starting from the initial state, we  can finish LTL task w.p.1 under some policy.  If not, we can directly output ``no solution''.
  
\section{Solution for Communicating MDP} \label{sec:4}
In this section, 
instead of tackling the general case, 
we focus on solving Problem~\ref{problem:maxicons} for \emph{communicating MDPs}, i.e., 
MDPs in which all states can visit each other under some policy.  
Formally, an MDP $\mathcal{M}$ is said to be \emph{communicating} if
\[
\forall s,s'\in S, \exists  \mu \in \Pi_{\mathcal{M}}^S,\exists n\geq 0:  (\mathbb{P}^{\mu})^{n}_{s,s'}>0.
\]
In fact, if $\M$ is communicating, then the above condition can be achieved by a stationary policy $\mu \in \Pi^{S}_\mathcal{M}$.  
Therefore, the resulting MC $\M^{\mu}$ is \emph{irreducible}, i.e., each pair of states can be visited from one to the other via some path. 

 Note that, without considering the LTL task, 
the computation of the entropy rate for a general MDP has been addressed in \cite{LIPIcs:2014:4872}. 
However, the  approach in \cite{LIPIcs:2014:4872} does not directly yield a policy to achieve this value. 
Here, instead of considering the general MDPs, 
we first focus only on communicating MDPs, and show that, for this class of MDPs, 
the stationary policy achieving entropy rate maximization can be synthesized more efficiently based on a different nonlinear program \eqref{opt1:obj}-\eqref{opt1:con5} which describes a steady-state parameter synthesis problem. The intuition of this nonlinear program is as follows. 
Equation~\eqref{opt1:con5} indicates that the decision variables here are $\gamma(s,a)$ for each state-action pair $s \in S$ and  $a \in A(s)$.
They are used to represent the probability of occupying state $s$ and choosing action $a$ when the system goes to steady state, i.e., $\sum_{a \in A(s)}\gamma(s,a)=\pi(s)$ with $\pi$ the limit distribution of the induced MC.  
Variables $q(s,t)$ and $\lambda(s)$ in Equations~\eqref{opt1:con1} and~\eqref{opt1:con2} are functions of $\gamma(s,a)$, 
representing the probability of going from states $s$ to $t$ and the probability of occupying state $s$, respectively.  
Equations~\eqref{opt1:con3} and~\eqref{opt1:con4} are constraints for stationary distribution that the decision variables should satisfy.  
Finally, the objective function is determined according to the computation of entropy rate  given in Equation~\eqref{(8)}.
 \begin{tcolorboxenv}
\begin{center}
        \tcbset{title=\quad Nonlinear Program for Communicating MDP,colback=white}
    \begin{tcolorbox} 
    \vspace{-5pt}
\begin{align}
 &\max_{\gamma(s,a)} \quad \sum_{s \in S} \sum_{t \in S} -q(s,t)\log \left(\frac{q(s,t)}{\lambda(s)}\right)   \label{opt1:obj} \\
\!\!\!\!\!\!\!\!\text{s.t. } \ \
&q(s,t) = \sum_{a \in A(s)} \gamma(s,a)P(t \mid s,a),\forall s, t \in S   \label{opt1:con1}\\
&\lambda(s) = \sum_{a \in A(s)} \gamma(s,a), \forall s \in S  \label{opt1:con2}\\    
&\lambda(t) = \sum_{s \in S} q(s,t),\forall t \in S  \label{opt1:con3} \\
&\sum_{s \in S} \lambda(s)=1     \label{opt1:con4}\\
&\gamma(s,a) \geq 0, \forall s \in S ,\forall a \in A(s)   \label{opt1:con5}
\end{align} 
    \end{tcolorbox}
\end{center}
\end{tcolorboxenv}
Now, given a communicating MDP $\M$, let $\gamma^{\star}(s,a)$ be the solution to  Equations~\eqref{opt1:obj}-\eqref{opt1:con5}. 
Then the maximum entropy rate is achieved by the stationary policy defined as
\begin{equation}\label{eq:opt-sta}
    \mu^{\star}(s,a)=\frac{\gamma^{\star}(s,a)}{\sum_{a \in A(s)}\gamma^{\star}(s,a)}, \quad \forall s \in S, \forall a \in A(s).
\end{equation}
We now prove that Equation~(\ref{eq:opt-sta}) simply decodes a policy that achieves the best stationary distribution of the desired MC.
\begin{mypro} \label{pro:optimalCMDP}
Let $\mathcal{M}$ be a communicating MDP. 
Then for policy $\mu^{\star} \in \Pi^{S}_\mathcal{M}$ defined by \eqref{eq:opt-sta}, we have $\nabla H(\M^{\mu^{\star}})=\nabla H(\mathcal{M})$.
\end{mypro}
\begin{proof}
The proof is provided in the Appendix. 
\end{proof}
\begin{remark}[\bf Complexity of the Nonlinear Program] 
Here we would like to remark that the proposed nonlinear program in  Equations~\eqref{opt1:obj}-\eqref{opt1:con5} is convex. 
Therefore, it can be solved in polynomial-time. 
First, we note that constraints~\eqref{opt1:con1}-\eqref{opt1:con5} are affine.  
Furthermore, we note that the objective function~\eqref{opt1:obj} is concave when $\lambda(s,a) \geq 0$. 
To see this, for $s,t \in S$, we define $\ell(s,t)=q(s,t)\log(\frac{q(s,t)}{\lambda(s)})$, i.e., $\ell(s,t)$ is the relative entropy and it is convex w.r.t. $(q(s,t), \lambda(s))$ \cite[Page 90]{boyd2004convex}. Then the objective function is equal to $-\sum_{s \in S} \sum_{t \in S} \ell(s,t)$ and it is concave w.r.t.   vector $(q(s,t),\lambda(s))_{s \in S, t\in S} \in \mathbb{R}^{|S|^2+|S|}$. From \cite[Page 79]{boyd2004convex}, the composition of concave function with an affine mapping is also a concave function. Since  vector $(q(s,t),\lambda(s))_{s \in S, t\in S}$ is affine in vector $(\gamma(s,a))_{s\in S, a\in A(s)}$, $-\sum_{s \in S} \sum_{t \in S} \ell(s,t)$ is concave w.r.t. vector $(\gamma(s,a))_{s\in S, a\in A(s)}$.
Therefore, a solution of (\ref{opt1:con1})-(\ref{opt1:con5}) whose objective value (\ref{opt1:obj}) is arbitrarily close to the optimal value can be computed in polynomial-time 
in the size of $\M$.  
\end{remark}  
 
Note that, the above nonlinear program does not take the satisfaction of the LTL task into account. 
In order to satisfy the LTL task, 
one needs to stay in some AMEC forever. To handle it, first, we show the following structural property for communicating MDPs. 
It reveals that, for a communicating MDP, if a stationary policy maximizes the entropy rate, then its induced MC must be also irreducible.  
 
\begin{mylem} \label{prop:com_irr}
Let $\mathcal{M}$ be a communicating MDP. From \cite{LIPIcs:2014:4872} there exists stationary policy $\mu \in \Pi^S_\M$ such that $\nabla H(\M)=\nabla H(\M^\mu)$. Then the induced MC $\mathcal{M}^{\mu}$ is irreducible.
\end{mylem}
\begin{proof}
The proof is provided in the Appendix. 
\end{proof}
Recall that, if an MC is irreducible, then the limit distribution is initial-distribution independent~\cite{puterman}. 
Therefore, the above structural property implies that   $\nabla H(\M)$ is also initial-distribution independent. 
Note that a sub-MDP $(\S,\A) \in \texttt{MEC}(\M)$ is  always communicating. 
In rest of paper, we will write the maximum entropy rate of sub-MDP $(\S,\A)$ by $\nabla H(\S,\A)$ directly by ignoring the initial distribution. To solve the Problem~\ref{problem:maxicons} for communicating MDP, we first compute all AMECs 
and for each AMEC, which is also a communicating MDP, 
we compute the optimal policy that maximizes the entropy rate within this AMEC. 
Then the overall strategy is to ensure that one can reach the AMEC with maximum entropy rate w.p.1. Note that the structural property is essential to guarantee the correctness of approach above. Specifically, by the structural property, the maximum entropy rate policy over AMEC will visit all states in it infinitely often. Then from the definition of AMEC, we know that this policy can also finish LTL task w.p.1. The idea above is summarized by Algorithm~\ref{alg:cmdpsolution}, 
where lines 12-16 is a procedure \cite[Page 480]{puterman} ensuring that the AMEC with maximum entropy rate can be reached eventually w.p.1. Note that line 13 is always feasible, i.e., there always exists $s \in T,a \in A(s)$ such that $\sum_{t \in G} P_{s,a,t}>0$. Otherwise it violates the assumption that the MDP $\mathcal{M}$ is communicating.

\begin{algorithm}[htp]
\caption{Solution for Communicating MDP}\label{alg:cmdpsolution} 
\KwIn{communicating MDP $\M= (S,s_0,A,P,\mathcal{AP},\ell,Acc)$}
\KwOut{optimal policy $\mu^\star \in \Pi^{S}_{\M}$ and its associated maximum entropy rate $v^\star$ of MC $\M^{\mu^\star}$} 

compute $\texttt{AMEC}(\M)=\{ (\S_1,\A_1),\dots,(\S_n,\A_n) \}$\\

\If{$\texttt{AMEC}(\M)=\emptyset$}
{
$\mu^\star\gets$ arbitrary policy,  and $v^\star\gets -\infty $
}
\Else
{
\For{\text{each sub-MDP}  $(\S_i,\A_i),i=1,\dots,n$}
{
compute maximum entropy rate policy $\mu_i$  by  Equation~\eqref{eq:opt-sta} 
and the associated value $v_i=\nabla H(\S_i,\A_i)$ 
}
$i^\star\gets \arg\max_{i}\{ v_1,v_2,\dots,v_n\}$ \\

for $s \in \S_{i^\star}$, $a \in \A_{i^\star}(s)$,  $\mu^\star(s,a)\gets \mu_{i^\star}(s,a)$  \\
 $T\gets S\setminus \S_{i^\star}$, and  $G\gets \S_{i^\star} $ \\
\While{$T \neq \emptyset$}
{
pick $s \in T,a \in A(s)$ \text{ s.t. } $\sum_{t \in G} P_{s,a,t}>0$ \\
 $\mu^\star(s,a)\gets 1$ \\
$T \gets T\setminus \{ s\}$, and $G\gets G \cup \{s \}$
}
}
\end{algorithm}

The following result shows that, for a communicating MDP,  
the policy computed by Algorithm~\ref{alg:cmdpsolution} indeed (i) satisfies the LTL task; and (ii) maximizes the entropy rate. 
\begin{mythm} \label{thm:alg1isright}
Given MDP $\M=(S,s_0,A,P,\mathcal{AP},\ell,Acc)$ that is communicating, the output policy of Algorithm~\ref{alg:cmdpsolution} is a solution to Problem~\ref{problem:maxicons} for $\M$.
\end{mythm}
\begin{proof} 
According to Lemma~\ref{prop:com_irr}, for each AMEC $(\S_i,\A_i)$, policy $\mu_i$ ensures that all states in it will be visited infinitely often w.p.1. 
Based on the action assignment procedure in lines 12-16, MC $\M^{\mu^\star}$ has only one recurrent class $\S_i$. 
By the definition of AMEC, we have $\mu^\star \in   \Pi^{\varphi}_{\M}$, i.e.,  $\nabla H(\M^{\mu^\star}) \leq \nabla H_{\varphi}(\M)$.

We now consider arbitrary $\mu \in \Pi^{S}_{\M} \cap   \Pi^{\varphi}_{\M}$. 
Let $R_1,R_2,\dots,R_K \subseteq S$ be the recurrent classes in $\M^{\mu}$. 
Since $\mu \in \Pi^{\varphi}_{\M}$, for any $R_k$, we can find $(\S,\A) \in \texttt{AMEC}(\M)$ such that $R_k \subseteq \S$. 
Let $r_k$ be the entropy rate restricted on recurrent class $R_k$. 
We denote by $r_k^\star$ maximum entropy rate of the AMEC that $R_k$ belongs to. 
Then by Claim~1 in the Appendix, we can find a set of values $\beta_1,\dots\beta_K\in[0,1]$ with $\sum_{k=1}^K\beta_k=1$ such that 
\[
\nabla H(\M^\mu) = \sum_{k=1}^{K} \beta_{k} r_k \leq \sum_{k=1}^{K} \beta_{k} r^\star_k \leq v_{i^\star} = \nabla H(\M^{\mu^\star}),
\]
where the first equality and the last inequality come from Claim~\ref{prop:limitdistri} and line 9 of Algorithm~\ref{alg:cmdpsolution}, respectively. 
Thus
\[
\nabla H(\M^{\mu^\star}) \geq \sup_{\mu \in \Pi^{S}_{\M} \cap   \Pi^{\varphi}_{\M}} \nabla H(\M^{\mu}) = \nabla H_\varphi(\M), 
\]
where the last equality comes from Proposition~\ref{prop:suffisenouth}.  Hence, we have $\nabla H(\M^{\mu^\star})=\nabla H_\varphi(\M)$. 
\end{proof}

\section{State Level  Classification}\label{sec:5}
\begin{figure}
    \centering\includegraphics[width=5cm]{ 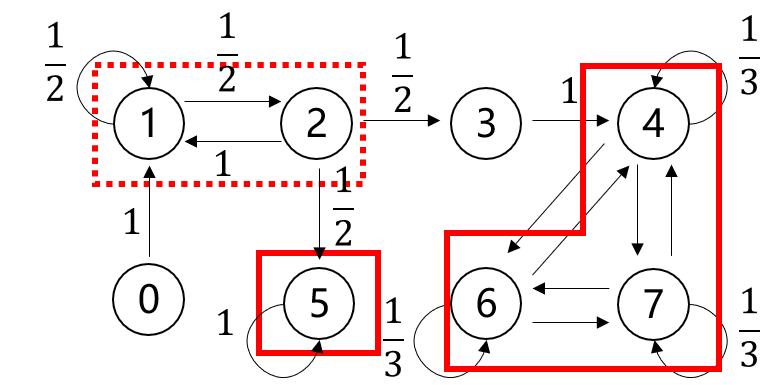}
 
    \caption{Illustrative example of state level classification. The state $0$ is the initial state. All states except for state $2$ have only one available action and thus we omit the action notation. The transition probability is illustrated on the edge. We also omit some transition probabilities among states $4,6$ and $7$ which are all equal to $\frac{1}{3}$. The state $2$ has two available actions. One action will reach state $1$ with probability $1$. The other action will reach states $5$ and $3$  with probability $0.5$ respectively.}
    \label{fig:example1}
\end{figure}

In the previous section, we have solved Problem~\ref{problem:maxicons} for the case of communicating MDPs. 
However, for the general case of non-communicating MDPs, this problem is much more challenging. 
Particularly, to enforce the satisfaction of the LTL task, the MDP should eventually stay in some MECs and different MECs may result in different entropy rates. Moreover, we cannot arbitrarily choose the AMEC to stay in forever with probability $1$ since there is no communicating assumption anymore.
For example, consider the MDP in Figure~\ref{fig:example1} with initial state $0$. Different from simple communicating MDP, for this multi-chain general MDP, there is no policy to reach any of the MECs in solid framess with probability $1$. To resolve this issue, this section proposes an approach to classify MECs into different ``levels'' in terms of their connectivities.
\subsection{Definitions of State Levels} 

Let 
$\texttt{MEC}(\mathcal{M})=\{ (\S_1,\A_1), \dots, (\S_n,\A_n) \}$
be the set of all MECs of MDP $\mathcal{M}$.  
For each state $s\in S$, we say it is 
\begin{itemize}
  \item 
  an \emph{MEC state} if it belongs to some MEC, and we denote by  $\S_M=\bigcup_{i=1}^n \S_i$ the set of MEC states; and 
  \item 
  a \emph{transient state} if it does not belong to any MEC, and we denote by 
  $\T=S\setminus \S_M$ the set of  all transient states. 
\end{itemize}

Note that each state can only belong to at most one MEC. 
Then for each MEC state $s\in \S_M$, we denote by 
$(\S_{[s]},\A_{[s]})$   the unique MEC it belongs to.
Also, we note that,   
for two different MECs 
$(\S_i,\A_i)$ and $(\S_j,\A_j)$, if $\S_i$ is reachable from $\S_{j}$, 
then $\S_j$ must be not reachable from $\S_i$; 
otherwise $(\S_i\cup \S_j,\A_i\cup \A_j)$ will be a larger MEC. 
Based on this observation,  we can classify MECs into different ``levels'' as follows.
\begin{itemize}
\item 
First, there must exist states whose reachable sets remain entirely contained within  their MECs  and we consider those MEC states as the ``lowest'' level.
\item 
Then, for any MEC state, it is said to be with level $k$ if it can only reach MECs with lower levels  or  itself.  
\end{itemize}
The above discussion leads to the following definition.

\begin{mydef}[\bf State Levels for MEC States]
Let $\M$ be an MDP and $S_M$ be the set of MECs states. 
Then for each $k\geq 0$, the set of  $k$-level MEC states, denoted by $L_k$, is defined inductively as follows:
\begin{align} \label{eq:MEClevel}
    L_{0} \!=& \{ s \!\in \!\S_{M} \mid \textsf{reach}(s) = \S_{[s]} \}, \\
L_{k}\!=&\{ s \!\in \!\S_{M}  \mid \textsf{reach}(s) \!\cap\! \S_{M} \!   \subseteq\!  \bigcup_{m<k}\!  \ L_{m} \!\cup\! \S_{[s]} \}\setminus\bigcup_{m<k}\! L_{m}. \nonumber
\end{align}
We define $\texttt{level}(\M)=\max \{ k\mid L_k\neq \emptyset      \}$ 
as the highest level of MECs in MDP $\M$. 
\end{mydef}

Similarly, for transient states in $\T=S\setminus \S_M$, we also define the set of $k$-level states as collection of those states which can only reach MECs with levels smaller than or equal to $k$.  
\begin{mydef}[\bf State Levels for Transient States]
Let $\M$ be an MDP and $\T=S\setminus \S_M$ be the set of transient states.  
Then for each $k\geq 0$, the set of  $k$-level transient states, denoted by $T_k$, is defined by:
\begin{equation} \label{eq:tranlevel}
T_{k}=\{ s \in  \T   \mid \textsf{reach}(s) \cap \S_{M} \subseteq \bigcup_{m\leq k} L_{m} \}\setminus \bigcup_{m<k} T_{m}. 
\end{equation}
\end{mydef}
Note that for a transient state, it can always reach some MEC state, i.e.,  $\forall s \in \mathcal{T}, \textsf{reach}(s)\cap \mathcal{S}_M \neq \emptyset$. 
Also, the state levels for transient state is defined based on state levels of MEC states. Specifically, if $s \in T_k$, then $s$ can reach some MEC in $L_k$ and cannot reach MECs with higher level. Moreover,  $s \in L_k$ cannot reach $t \in T_k$; otherwise, $s$ can reach other MEC in $L_k$ through $t$, which contradicts to the definition of MEC.
We use the example in Figure~\ref{fig:example1} to illustrate the definitions.

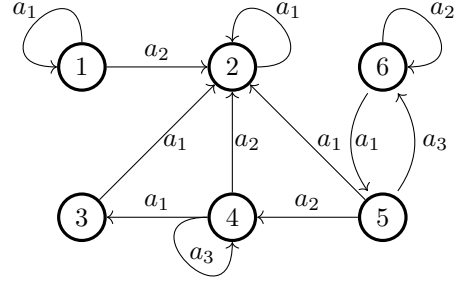
\begin{figure}
    \centering
    \begin{tikzpicture}
[
square/.style={circle, draw=black!255, fill=white!255, very thick, minimum height=5mm,minimum width=5mm},
]
	\node [square](q1)at(0,2){$1$};
	\node [square](q2)at(2,2){$2$};
	\node [square](q3)at(0,0){${3}$};
	\node [square](q4)at(2,0){${4}$};
	\node [square](q5)at(4,0){${5}$};
 	\node [square](q6)at(4,2){${6}$};
	\node (a12)at(1,2.2){$a_{2}$};
	\node (a22)at(2.75,2.75){$a_{1}$};
 \node (a32)at(1.25,1){$a_{1}$};
	\node (a11)at(-0.75,2.75){$a_{1}$};
	\node (a42)at(2.2,1){$a_{2}$};
	\node (a43)at(1,0.2){$a_{1}$};
     \node (a44)at(1.6,-0.5){$a_{3}$};
         \node (a52)at(3.3,1){$a_{1}$};
         \node (a54)at(3,0.2){$a_{2}$};
	\node (a56)at(4.7,1){$a_{3}$};
	\node (a65)at(3.8,1){$a_{1}$};
	\node (a66)at(4.8,2.75){$a_{2}$};
	\draw[->] (q1) -- (q2);
	\draw[->] (q3) -- (q2);
	\draw[->] (q4) -- (q2);
	\draw[->] (q4) -- (q3);
        \draw[->] (q5) -- (q2);
        \draw[->] (q5) -- (q4);
 	\draw[->] (4.2,0.35)..controls (4.5,0.78) and (4.5,1.21) .. (4.2,1.65);
   	\draw[->] (3.8,1.65)..controls (3.5,1.21) and (3.5,0.78) .. (3.8,0.35);
	\draw[->] (q2) .. controls +(right:15mm) and +(up:15mm) .. (q2);
	\draw[->] (q1) .. controls +(up:15mm) and +(left:15mm) .. (q1);
	\draw[->] (q6) .. controls +(up:15mm) and +(right:15mm) .. (q6);
 	\draw[->] (q4) .. controls +(left:15mm) and +(down:15mm) .. (q4);
    \end{tikzpicture}\vspace{-20pt} 
    \caption{Illustrative example of state level classification.}
    \label{fig:example}
\end{figure}

\begin{myexm}
Let us consider an MDP $\M$ shown in Figure~\ref{fig:example1}. 
This MDP has three MECs $\{ (\S_1,\A_1), (\S_2,\A_2),(\S_3,\A_3) \}$, 
where $\S_1=\{ 1,2 \}$ with $\A_1(1)=\A_1(2)=\{ a_1 \}$,
$\S_2=\{ 5 \}$ with $\A_2(2)=\{ a_1 \}$,
$\S_3=\{ 4, 6, 7 \}$ with $\A_3(4)=\A_3(6)=\A_3(7)=\{ a_1 \}$. 
Then we have  $\S_M = \{ 1,2,4,5,6,7 \}$ and $\mathcal{T}=\{0, 3\}$. 
Clearly,  $L_0=\{ 4,5,6,7 \}$ since $\textsf{reach}(i)\cap \S_M=\S_{[i]}$ for $i\in \{ 4,5,6,7 \}$. 
For $i \in \{ 1,2,4,5,6,7 \}$, since $\textsf{reach}(i) \cap \S_M \subseteq L_0 \cup \S_{[i]}$,  we have $L_1=\{ 1,2,4,5,6,7 \}\setminus L_0 = \{ 1,2 \}$.  
Since $\textsf{reach}(3)\cap \S_M=\{4,5,6,7 \} \cap \S_M \subseteq L_0$, we have $T_0=\{3 \}$. Similarly, we have $T_1=\{ 0 \} $. 
Finally, we have $\texttt{level}(\mathcal{M})=1$.
\end{myexm}
Another illustrative example is shown as below.
\begin{myexm}
Let us consider an MDP $\M$ shown in Figure~\ref{fig:example}.
For each action, the transition probability is one and the value is omitted in the figure. 
This MDP has four  MECs $\{ (\S_1,\A_1), (\S_2,\A_2),(\S_3,\A_3),(\S_4,\A_4) \}$, 
where $\S_1=\{ 1 \}$ with $\A_1(1)=\{ a_1 \}$,
$\S_2=\{ 2 \}$ with $\A_2(2)=\{ a_1 \}$,
$\S_3=\{ 4 \}$ with $\A_3(4)=\{ a_3 \}$, and 
$\S_4=\{ 5,6 \}$ with $\A_4(5)=\{ a_3 \}, \A_4(6)=\{a_1,a_2\}$. 
Then we have  $\S_M = \{ 1,2,4,5,6 \}$ and $\mathcal{T}=\{ 3\}$. 
Clearly,  $L_0=\{ 2 \}$ since $\textsf{reach}(2)\cap \S_M=\S_{[2]}$. 
For states    $i=1,2,4$, since $\textsf{reach}(i) \cap \S_M \subseteq L_0 \cup \S_{[i]}$,  we have $L_1=\{ 1,2,4 \}\setminus L_0 = \{ 1,4 \}$. 
Similarly, we have  $L_2=\{ 5,6 \}$. 
Since $\textsf{reach}(3)\cap \S_M=\{2 \} \cap \S_M \subseteq L_0$, we have $T_0=\{3 \}$ and $T_1=T_2=\emptyset$. 
The level of the MDP is $\texttt{level}(\mathcal{M})=2$.
\end{myexm}

\begin{remark}
  The proposed state-level classification method holds its own value beyond our problem. Unlike classical MEC computation algorithms, which only identify MECs, our method additionally
reveals the connectivity hierarchy among them. This structural insight could benefit other problems, for instance, in
model checking or reward maximization, where transition probabilities frequently change. By leveraging state-level
information, recomputation can be localized to only those states with higher levels than the modified ones, potentially
yielding computational savings in practice.
\end{remark}

\subsection{Computations of State Levels} 
In order to classify the level of each state,   
we first define a relation 
$  \leq_L \subseteq \texttt{MEC}(\M) \times \texttt{MEC}(\M)$  
by:  for any $U_1=(\S_1,\A_1),U_2=(\S_2,\A_2)\in \texttt{MEC}(\M)$, we have 
\begin{equation} \label{eq:partialdefi}
    U_1 \leq_L U_2 \iff   \forall s \in \S_2: \S_1 \subseteq \textsf{reach}(s).
\end{equation}
\begin{mypro}
Relation  
$  \leq_L \subseteq \texttt{MEC}(\M) \times \texttt{MEC}(\M)$ defined in Equation~\eqref{eq:partialdefi}  is a partial order relation.
\end{mypro}
\begin{proof}
We show that $\leq_L$ is reflexive, anti-symmetric and transitive. 
First, by the definition of MEC, we have $\S_{[s]} \subseteq \textsf{reach}(s)$. 
Thus $(\S,\A) \leq_L (\S,\A)$, i.e., $\leq_L$ is reflexive. 
Second, for two MECs $(\S_1,\A_1) \neq (\S_2,\A_2)$, if $(\S_1,\A_1) \leq_L (\S_2,\A_2)$ and $(\S_2,\A_2) \leq_L (\S_1,\A_1)$, then $(\S_1 \cup \S_2, \A_1 \cup \A_2)$ is also an MEC. However, this will  contradict to the fact that $(\S_1,\A_1)$ and $(\S_2,\A_2)$ are two different MECs. 
Thus, $\leq_L$ is anti-symmetric. 
Finally, If $(\S_1,\A_1) \leq_L (\S_2,\A_2)$ and $(\S_2,\A_2) \leq_L (\S_3,\A_3)$, then for any $s \in \S_3$, $\S_2 \subseteq \textsf{reach}(s)$ and for any $t \in S_2$, $\S_1 \subseteq \textsf{reach}(t)$. 
Thus $\S_1 \subseteq \textsf{reach}(s)$, i.e., $(\S_1,\A_1) \leq_L (\S_3,\A_3)$, which means that $\leq_L$ is transitive.   
\end{proof}
Note that partial order $\leq_L$ can be computed directly by checking the reachability between MECs.
With partial order $\leq_L$,  we can compute state levels for MEC states by Algorithm~\ref{alg:classification}. 
Specifically,  since $\texttt{MEC}(\M)$ is a finite set, there must exist minimal elements in relation $\leq_L$, and $L_0$ contains all states in MECs that are minimal elements in $\leq_L$. Then we eliminate these elements in $\leq_L$ and the minimal elements for remaining relation consist the $L_1$. 
We repeat this procedure and find MECs of each level. 
The level for each transient state $s \in \mathcal{T}$ is computed in lines 11-14, which is simply the highest level of MEC states it can reach.
\begin{algorithm} 
\caption{State Level Classification}\label{alg:classification} 
\KwIn{MDP $\M= (S,s_0,A,P,\mathcal{AP},\ell,Acc)$}
\KwOut{$L_0',T_0',\dots,L_{\texttt{level}(\M)}',T_{\texttt{level}(\M)}'$\vspace{3pt}}

compute  $\texttt{MEC}(\M)=\{ (\S_1,\A_1),\dots,(\S_n,\A_n) \}$\\
compute partial order $\leq_L$ in \eqref{eq:partialdefi} \\
$num\gets 0$ \\
\While{$\leq_L \neq \emptyset$}
{
$\mathcal{I}  \gets  \{ i \mid (\S_i,\A_i) \text{ is minimal element in}\leq_L \}$\\
$L_{num}'\gets   \bigcup_{ i \in \mathcal{I}} \S_i$ \\
$\leq_L \gets  \leq_L \setminus \{ ((\S_i,\A_i),(\S_j,\A_j)) \mid i \in \mathcal{I} \}$ \\
$num\gets num+1$ 
}
$T_0'=T_1'=\dots=T_{num-1}' \gets \emptyset$ \\
\For{$s \in S \setminus \bigcup_{i=1}^{k}\S_i$}
{
$cnt\gets\max\{ j \mid \textsf{reach}(s) \cap L_j' \neq \emptyset \}$\\
$T_{cnt}'\gets T_{cnt}'\cup \{ s \}$
}
\end{algorithm}
\begin{mypro} \label{prop:classisright}
Given MDP $\M= (S,s_0,A,P,\mathcal{AP},\ell,Acc)$, the outputs of Algorithm~\ref{alg:classification} are indeed $L_0,L_1,\dots,L_{\texttt{level}(\M)}$ defined in Equation~(\ref{eq:MEClevel}) and $T_0,T_1,\dots,T_{\texttt{level}(\M)}$ defined in Equation~(\ref{eq:tranlevel}).
\end{mypro} 
\begin{proof} 
We first prove by induction that, at $n$-th time the algorithm executes the ``while''-loop in lines 4-9, we have $L_{n-1}'=L_{n-1}$. 
When $n=0$, since $\leq_L$ is finite, we know that there exists minimum elements in $\leq_L$.  
If $i \in \mathcal{I}$, we know that for $s \in \S_i$, $\textsf{reach}(s)\cap \S_M = \S_i$. Thus $L_0'=L_0$. Assume that $L_0'=L_0,\dots,L_k'=L_k$ for some $k\geq 0$ and $\leq_L\neq \emptyset$ after eliminating the minimum elements in each ``while''-loop. For $n=k+1$, if $i \in \mathcal{I}$, it means that for all $s \in \S_i$,
\[
\textsf{reach}(s)\cap \S_M \subseteq \bigcup_{m < k+1}L'_m \cup \S_{[s]}= \bigcup_{m < k+1}L_m \cup \S_{[s]},
\]
where last equality comes from induction hypothesis. Since $L'_{k+1} \cap (\bigcup_{m<k+1} L'_{m})=\emptyset$, from \eqref{eq:MEClevel} it holds that $L'_{k+1}=L_{k+1}$. The proof of induction is completed.

Next, we show that $T_0',T_1',\dots,T_{num-1}'$ are indeed $T_0, T_1,$ $\dots,T_{\texttt{level}(\M)}$. 
In above, we have $\texttt{level}(\M)=num-1$. 
For $T_k$ defined in (\ref{eq:tranlevel}), $s \in T_k $ if and only if $s \in \mathcal{T}$ and
\[
[L_k \cap \textsf{reach}(s) \neq \emptyset] \wedge (\forall i > k)[L_i \cap \textsf{reach}(s) = \emptyset].
\]
Therefore, $T_k'$ exactly matches the definition of $T_k$.
\end{proof}

\section{Solution for General Case}\label{sec:6} 
In this section, we tackle the Problem~\ref{problem:maxicons} for general MDPs
based on the solutions for communicating MDPs and the state level classification technique in the previous two sections. 

\subsection{General Ideas}
For non-communicating MDP $\mathcal{M}$, we define
\[
R_k=\bigcup_{m=0}^{k}(L_{m}\cup T_{m})\text{ and } A_{k} : R_k \to 2^A \text{ s.t. } A_{k}(s)=A(s)
\]
as the set of all MEC states and transient states whose levels  are smaller than or equal to $k$, and the associated available actions, respectively.   
Similarly, we also define 
\[
\hat{R}_k=R_k\setminus T_k\text{ and } \hat{A}_{k}: \hat{R}_k \to 2^A\text{ s.t. }\hat{A}_{k}(s)=A(s)
\]
as the set of all MEC states with levels smaller than or equal to $k$ and transient states with levels strictly smaller than $k$, and  the associated available actions, respectively.  

Now we make the following observations for the above defined $R_k$ and $\hat{R}_k$. 
First, we observe that, for any $k=0,1,\dots,\texttt{level}(\M)$, 
$(R_k,A_{k})$ and $(\hat{R}_k,\hat{A}_{k})$ are both sub-MDPs. 
This is because, by the definition of state levels, states with level $k$ can only go to states with lower levels. Moreover, states in $L_k$ can never reach states in $T_k$ as discussed in  the definition of state levels for transient states.
Second, for sub-MDPs $(\hat{R}_k,\hat{A}_{k})$ and $(\hat{R}_m,\hat{A}_{m})$, where 
$k<m$, 
suppose that $\mu_k$ and $\mu_m$ are policies that solve Problem~\ref{problem:maxicons} for $(\hat{R}_k,\hat{A}_{k})$ and $(\hat{R}_m,\hat{A}_{m})$, respectively.  
By modifying $\mu_m$ to $\mu_m'$ such that $\mu_m(s,a)$ is
(i) changed to $\mu_k(s,a)$, for all $s\in \hat{R}_k$, 
and (ii) unchanged otherwise, we know that the modified $\mu_m'$  also solves Problem~\ref{problem:maxicons} w.r.t. $(\hat{R}_m,\hat{A}_{m})$.

The above observations suggest that we can find a solution to Problem~\ref{problem:maxicons}  in a backwards manner from states with the lowest level as follows:

\textbf{Step 1: }Initially, we start from those MECs  with level $0$ and compute  solutions for these sub-MDPs. 
Since each MEC is communicating, we can use Algorithm~\ref{alg:cmdpsolution} in Section~\ref{sec:4}.

\textbf{Step 2: }
Once all MECs in $L_0$ have been processed, we move to include transient states in $T_0$. 
This   provides an instance of Problem~1 w.r.t.\ sub-MDP $(R_0,A_0)$.  
Since states in $T_0$ are transient no matter what actions we take, the only factor that determines the total entropy rate is what MECs in $L_0$ they choose to go. 
Therefore, it suffices to solve an \emph{expected total reward} maximization problem, where the reward of reaching each MEC in $L_0$  is the computed maximum entropy rate.   

\textbf{Step 3: }
Then we  proceed to further consider MECs in $L_1$ in addition to $(R_0,A_0)$,  which gives an instance of Problem~1 w.r.t.\ sub-MDP $(\hat{R}_1,\hat{A}_1)$.   
Still, for each MEC in $L_1$, we   use Algorithm~\ref{alg:cmdpsolution} to compute the maximum entropy rate
within it.
However, here we have two  alternatives  for each MEC in $L_1$: 
\begin{enumerate}[(i)] 
    \item 
 consider it as a transient part as the case of $T_0$ by solving an expected total reward maximization problem; or
    \item 
    choose to stay in the current MEC forever.  
\end{enumerate}
Therefore, we need to compare these two alternatives and choose the one with larger reward (entropy rate).

\textbf{Step 4: }
Once $(\hat{R}_1,\hat{A}_1)$ is processed, we further include $T_1$ to consider the  instance of $({R}_1,{A}_1)$, and so forth, until the instance of $({R}_{\texttt{level}(\M)},{A}_{\texttt{level}(\M)})=\M$ is solved.

\subsection{Synthesis Algorithm}

Now, we formalize the implementation details of the above idea. 
Suppose that,  at decision stage $k=0,\dots,\texttt{level}(\M)$,  we have the following information available: 
\begin{itemize}
    \item 
    the set of states have been processed: $\hat{R}_k\subseteq S$;  
    \item 
    the solution $\hat{\mu}_{k}\in \Pi^S_{\hat{\M}_k}$ to Problem~\ref{problem:maxicons} w.r.t. current sub-MDP $\hat{\M}_k=(\hat{R}_k,\hat{A}_k)$; 
    \item 
    the maximum entropy rate one can achieve from each state while satisfying the LTL task w.p.1, which is specified by a function $\texttt{val}_k:\hat{R}_k\to \mathbb{R}$. 
\end{itemize}
Then our objective is to find the optimal policy, denoted by  
$\hat{\mu}_{k+1}\in \Pi^S_{\hat{\M}_{k+1}}$, for a larger sub-MDP 
\[
\hat{\M}_{k+1}=(\hat{R}_{k+1}=\hat{R}_{k}\cup  T_k \cup L_{k+1},\hat{A}_{k+1}).
\] 
Our approach consists of the following four steps. 

\textbf{Optimal Transient-Enforcing Policy: }
Note that, although states in $T_k$ are always transient,  the system may choose to stay in $L_{k+1}$. 
Here, we first synthesize a  \emph{transient-enforcing policy}, denoted by $\hat{\mu}_{k+1}'$, 
such that all states in $L_{k+1}$ are enforced to be transient.  
This policy only applies to states in $T_k \cup L_{k+1}$. 
To this end, we first solve  a new linear program (LP)~(\ref{opt2:obj})-(\ref{opt2:con4}), 
where 
  $\alpha$ is an arbitrary distribution vector over $ T_k \cup L_{k+1}$
  such that all elements are non-zero, and 
  $\texttt{val} = \texttt{val}_k+\epsilon$  is reward vector over $\hat{R}_k$ 
  with $\epsilon > 0$ be an arbitrary  value ensuring that  $\texttt{val}(s)>0$ for any $s \in \hat{R}_k$.

 \begin{tcolorboxenv}
\begin{center} 
        \tcbset{title=\quad Linear Program for Transient-Enforcing Policy,colback=white}
    \begin{tcolorbox}
    \vspace{-5pt}
\begin{align}
&\max_{\gamma(s,a)} \quad  \sum_{s \in T_k \cup L_{k+1}} \sum_{t \in \hat{R}_{k}} \texttt{val}(t) \lambda(s,t)    \label{opt2:obj} \\
\!\!\!\!\!\!\!\!\text{s.t. } \   & \eta(s) - \sum_{t \in T_k \cup L_{k+1}} \lambda(t,s)  \leq \alpha(s),\forall s \in T_k \cup L_{k+1}     \label{opt2:con1}\\
&  \eta(s) = \sum_{a \in A(s)} \gamma(s,a), \forall s \in T_k \cup L_{k+1} & \label{opt2:con2}\\
&     \lambda(s,t) =\!\! \sum_{a \in A(s)} \!\! \gamma(s,a) P_{s,a,t},    \forall s \!\in \!T_k \cup L_{k+1}, t\! \in\! \hat{R}_{k+1}   \label{opt2:con3} \\
&   \gamma(s,a) \geq 0 ,\forall s \in T_k \cup L_{k+1},  \forall a \in A(s)  \label{opt2:con4} 
\end{align} 
\end{tcolorbox}
\end{center}
\end{tcolorboxenv}

Let $\gamma^\star$ be the optimal solution to LP (\ref{opt2:obj})-(\ref{opt2:con4}). 
Similar to the LP to the standard expected total reward maximization problem~\cite{puterman}, the optimal solution satisfies that, 
for each $s \in T_k \cup L_{k+1}$, there exists at most one action, denoted as  $a^\star(s) \in A(s)$ such that $ \gamma^\star(s,a^\star(s))>0$. 
Then we define deterministic transient-enforcing policy $\hat{\mu}_{k+1}'$ by 
\begin{align}\label{eq:trans-enf}
	\!\!{\hat{\mu}_{k+1}'}(s)\!=\! \!
		\left\{\!\!\!
		\begin{array}{ll}
			a^\star(s) & \! \text{if}\   \sum_{a \!\in\! A(s)}\!\!\gamma^\star(s,a) \!> \!0    \\
			\text{arbitrary}     & \! \text{if}\   \sum_{a \!\in\! A(s)}\!\!\gamma^\star(s,a) \!= \!0
		\end{array}
		\right.   
\end{align} 
where ${\hat{\mu}_{k+1}'}(s)$ denotes the unique action chosen w.p.1 at $s$.

Intuitively, decision variable  $\gamma(s,a)$ in  LP~(\ref{opt2:obj})-(\ref{opt2:con4}) is the 
 the expected number of visits to state $s$  and choosing action $a$ when the initial distribution is $\alpha$. Here, we require that all elements in $\alpha$ are non-zero in order to ensure that all states can be visited with non-zero probability. 
Variables $\eta(s)$ and $\lambda(s,t)$ in Equations (\ref{opt2:con2}) and (\ref{opt2:con3}) are functions of $\gamma(s,a)$, representing the expected number of visits to state $s$ and the expected number of transitions from $s$ to $t$, respectively. Equation (\ref{opt2:con1}) is the constraint of the probability flow. Finally, objective function in Equation~(\ref{opt2:obj}) multiplies the probability of reaching $\hat{R}_{k}$ and the reward (maximum entropy rate) in $\hat{R}_{k}$ and sums over $s \in T_k \cup L_{k+1}$, representing the entropy rate of corresponding policy under initial distribution $\alpha$.
Note that, here we add $\epsilon$ uniformly to the original reward $\texttt{val}_k$ in order to force the optimal policy to visit states in $\hat{R}_k$.

\textbf{Optimal Staying Policy: }
Note that the transient-enforcing policy $\hat{\mu}_{k+1}'$ may not be the optimal policy for states in $T_k \cup L_{k+1}$ since we force each state in $L_{k+1}$ to go to MECs with lower levels. 
However, states in $L_{k+1}$ can also choose to stay at level $k+1$. 
To capture this situation, for each 
MEC $(\S_i,\A_i)\in  \texttt{MEC}(\mathcal{M})$ with level $k+1$, i.e., $\S_i\subseteq L_{k+1}$, 
we denote  by
$\mu_{\text{stay},i}$ and $v_i^\star$ the outputs of Algorithm~\ref{alg:cmdpsolution} when considering $(\S_i,\A_i)$ as the input communicating MDP.   
Note that, since all MECs are disjoint, we can use a single policy, denoted by $\mu_{\text{stay}}$, as the optimal staying policy for each MEC. 

\textbf{Value Evaluations: }
To fuse the transient-enforcing policy and staying policy, we need to compute their value functions. 
Then for each state in $s\in \S_i$,  where $(\S_i,\A_i)$ is an MEC with level $k+1$, 
the \emph{stay value} of $s$ is the maximum entropy rate when $s$ stays in the MEC forever, i.e., 
\begin{align}\label{eq:stay}
	\texttt{v}_\text{stay}(s) = v_i^\star.
\end{align} 
To compute the value function under  transient-enforcing policy $\hat{\mu}_{k+1}'$, 
we define  matrix $\mathbb{P}^{\hat{\mu}_{k+1}'}_{\hat{R}_k} \in \mathbb{R}^{|T_k \cup L_{k+1}| \times |\hat{R}_k|}$ such that $\mathbb{P}^{\hat{\mu}_{k+1}'}_{\hat{R}_k}(s,t)=\sum_{a \in A(s)} \hat{\mu}_{k+1}'(s,a) P_{s,a,t}$ is the transition probability from $s \in T_k \cup L_{k+1}$ to $t \in \hat{R}_k$ under partial policy $\hat{\mu}_{k+1}'$. We can similarly define matrix  $\mathbb{P}^{\hat{\mu}_{k+1}'}_{T_k \cup L_{k+1}}$ computing transition probabilities between states in $T_k \cup L_{k+1}$.   
The \emph{transient value} function  $\texttt{v}_\text{trans} \in \mathbb{R}^{|T_k \cup L_{k+1}|}$ is the solution of following equation
\begin{equation}
\texttt{v}_\text{trans}= \mathbb{P}^{\hat{\mu}_{k+1}'}_{T_k \cup L_{k+1}} \texttt{v}_\text{trans}+ \mathbb{P}^{\hat{\mu}_{k+1}'}_{\hat{R}_{k}} \texttt{val}_{k}.
\end{equation}
Since $I-\mathbb{P}^{\hat{\mu}_{k+1}'}_{T_k \cup L_{k+1}}$ is invertible~\cite[page 595]{puterman}, we have
\begin{equation} \label{eq:nextval}
    \texttt{v}_\text{trans}=(I-\mathbb{P}^{\hat{\mu}_{k+1}'}_{T_k \cup L_{k+1}})^{-1}\mathbb{P}^{\hat{\mu}_{k+1}'}_{\hat{R}_{k}} \texttt{val}_{k}.
\end{equation}

\textbf{Policy Fusion: }
To fuse policies $\hat{\mu}_{k+1}'$, $\mu_{\text{stay}}$, and the low level policy $\hat{\mu}_{k}$, we consider three cases: 
\begin{itemize}
\item
\text{Case 1 :} $[s\in T_k]$ or $[s\in L_{k+1} \wedge \texttt{v}_\text{trans}(s)\!>\!\texttt{v}_\text{stay}(s)]$; 
\item
\text{Case 2 :} $[s\in L_{k+1} \wedge \texttt{v}_\text{trans}(s)\!\leq\!\texttt{v}_\text{stay}(s)]$;  
\item
\text{Case 3 :} $s\in \hat{R}_{k}$.
\end{itemize}
Then the  new stationary policy 
$\hat{\mu}_{k+1}$ is fused as follows:  
for each $s\in \hat{R}_{k+1}$, we have
\begin{align}\label{eq:fuse}
\hat{\mu}_{k+1}(s,a) = 
		\left\{
		\begin{array}{ll}
			\hat{\mu}_{k+1}'(s,a) &  \text{if } \quad  \text{Case 1}\\
			\mu_{\text{stay}}(s,a)      &  \text{if }  \quad  \text{Case 2}\\
			\hat{\mu}_{k}(s,a)   & \text{if } \quad  \text{Case 3}
		\end{array}.
		\right.   
\end{align} 
The value function for the fused policy $\hat{\mu}_{k+1}$ is updated as  
\begin{align}\label{eq:valk+1}
	\texttt{val}_{k+1}(s) = 
		\left\{
		\begin{array}{ll}
			\texttt{v}_\text{trans}(s) &   \text{if } \quad  \text{Case 1}\\
			\texttt{v}_\text{stay}(s)     &   \text{if }  \quad  \text{Case 2}\\
			\texttt{val}_{k}(s)   & \text{if } \quad  \text{Case 3}
		\end{array}.
		\right. 
\end{align}  
Note that, for $k=0$ and each $s\in \hat{R}_0$, we have 
\begin{equation}\label{eq:val0}
\texttt{val}_0(s)=v_{[s]}^\star\text{ and } \hat{\mu}_0(s,a)= \mu_{\text{stay},[s]}(s,a) 
\end{equation}
 where $v_{[s]}^\star$ and $\mu_{\text{stay},[s]}$ are maximum entropy rate and the corresponding optimal policy for staying within MEC $(\S_{[s]},\A_{[s]})$. 
Essentially, for each MEC, one should decide whether to stay in itself or reach other MECs with lower level to achieve higher entropy rate value. The complexity of directly enumerating all situations is exponential w.r.t. the number of MECs. Our state level-based backward optimization procedure leverages the connectivity information among MECs to reduce the complexity to be polynomial.

\begin{algorithm} 
\caption{Solution for General MDP}\label{alg:gmdpsolution} 
\KwIn{MDP $\M= (S,s_0,A,P,\mathcal{AP},\ell,Acc)$}
\KwOut{optimal policy $\mu^\star \in \Pi^{S}_{\M}$} 

compute $\texttt{MEC}(\M)=\{ (\S_1,\A_1),\dots,(\S_n,\A_n) \}$
\\
compute $\mu^\star_i$ and $v^\star_i$ for each $(\S_i,\A_i) \in \texttt{MEC}(\M)$ 
\\
classify states into $L_{0},T_{0,}\dots, L_{\texttt{level}(\mathcal{M})},T_{\texttt{level}(\mathcal{M})}$
\\
compute $\hat{\mu}_0$ and $\texttt{val}_0$ according to Eq.~\eqref{eq:val0}
\\ 
\For{$k=0,1,\dots,\texttt{level}(\M)$}
{
solve   LP (\ref{opt2:obj})-(\ref{opt2:con4}) for $(\hat{\M}_{k+1},\hat{R}_{k},\texttt{val}_{k})$
\\
compute policy $\hat{\mu}^{'}_{k+1}$ according to Eq.~(\ref{eq:trans-enf})
\\
compute value functions $\texttt{v}_{\text{stay}}$ and $\texttt{v}_{\text{trans}}$ 
according to Eq.~(\ref{eq:stay}) and  ~(\ref{eq:nextval}), respectively
\\
fuse  policy $\hat{\mu}_{k+1}$ according to Eq.~(\ref{eq:fuse}) and 
compute $\texttt{val}_{k+1}$ according to Eq.~(\ref{eq:valk+1})
}
\textbf{Return} $\hat{\mu}_{\texttt{level}(\mathcal{M})+1}$
\end{algorithm}
The overall synthesis procedure is summarized in Algorithm~\ref{alg:gmdpsolution}. 
First, we initialize the optimal policy as well as the value function for level $0$ in line~4. 
In lines 5-10, we iteratively compute the optimal policy $\hat{\mu}_k$ for each level $k$. 
When $k=\texttt{level}(\mathcal{M})+1$, $\hat{\mu}_{k}$ is already the optimal policy for the entire MDP. Note that sub-MDP $\hat{\M}_{\texttt{level}(\M)}$ has not included states in $T_{\texttt{level}(\M)}$ yet.  
\subsection{Correctness proof}
We conclude this section by analyzing the correctness of the synthesis algorithm. 
We start by proving that $\hat{\mu}'_{k+1}$ is indeed the optimal transient-enforcing policy. 
Let 
\begin{align}
&\Pi^{T}_{\hat{\M}_{k+1}}=\nonumber\\
&\{ \mu \in \Pi^{\varphi}_{\hat{\M}_{k+1}} \cap \Pi^{S}_{\hat{\M}_{k+1}} \mid T_{k}\cup L_{k+1} \text{ is transient in } \hat{\M}_{k+1}^{\mu}\}\nonumber
\end{align}
be the set of stationary policies under which all states in $T_{k} \cup L_{k+1}$ are transient and the LTL task is satisfied w.p.1.  
Then the following result shows that  policy $\hat{\mu}_{k+1}$ achieves maximum entropy rate among $\Pi^{T}_{\hat{\M}_{k+1}}$.
\begin{mypro} \label{prop:transientoptimal}
Let  $\hat{\mu}_k$ be a solution to Problem~\ref{problem:maxicons} w.r.t. MDP $\hat{\M}_k=(\hat{R}_k,\hat{A}_k)$ regardless of the initial state 
and $\hat{\mu}_{k+1}'$ be the policy defined in \eqref{eq:trans-enf}. 
We define policy $\tilde{\mu}_{k+1}$ by: $\tilde{\mu}_{k+1}(s,a)=\hat{\mu}_{k+1}'(s,a)$ for $s \in T_k \cup L_{k+1}$ and $\tilde{\mu}_{k+1}(s,a)=\hat{\mu}_{k}(s,a)$ for $s \in \hat{R}_k$.  
Then regardless of the initial state of MDP $\hat{\M}_{k+1}=(\hat{R}_{k+1},\hat{A}_{k+1})$, we have
\[
\nabla H(\hat{\M}^{\tilde{\mu}_{k+1}}_{k+1}) = \sup_{\mu \in \Pi^{T}_{\hat{\M}_{k+1}} } \nabla H(\hat{\M}^{\mu}_{k+1}).
\]
Moreover, for $s \in T_k \cup L_{k+1}$, value $\texttt{v}_\text{trans}(s)$ in \eqref{eq:nextval} is the entropy rate $\nabla H(\hat{\M}^{\tilde{\mu}_{k+1}}_{k+1})$ when the initial state is $s$. 
\end{mypro}
\begin{proof}
The proof is provided in the Appendix. 
\end{proof}

Next, we prove that, if the initial state of MDP is an MEC state, then the maximum entropy rate  is the maximum of stay value and transient value.
\begin{mypro}\label{prop:MECmaximumisstayandtrans}
For MDP $\hat{\M}_{k+1}\!=\!(\hat{R}_{k}\!\cup\!  T_k \!\cup \!L_{k+1},\hat{A}_{k+1})$, suppose that the initial state satisfies $s_0 \in L_{k+1}$. Then
\begin{equation}\label{eq:stayandtransisenough}
    \nabla H_{\varphi}(\hat{\M}_{k+1})= \max\{ \texttt{v}_\text{stay}(s_0), \texttt{v}_\text{trans}(s_0) \},
\end{equation}
where $\texttt{v}_\text{stay}$ and $\texttt{v}_\text{trans}$ are defined in \eqref{eq:stay} and \eqref{eq:nextval}, respectively.
\end{mypro}
\begin{proof}
By \cite[Thm 8.3.2]{puterman}, for any two different initial states in $\S_{[s_0]}$, $\nabla H_\varphi(\hat{\M}_{k+1})$ are same. Thus, if $s_0 \in L_{k+1}$, we can find a solution $\tilde{\mu}_{k+1} \in \Pi^S_{\hat{\M}_{k+1}}$ of Problem~\ref{problem:maxicons} w.r.t. $\hat{\M}_{k+1}$ such that w.p.1, MC $\hat{\M}_{k+1}^{\tilde{\mu}_{k+1}}$ either 1) stays in MEC $(\S_{[s_0]},\A_{[s_0]})$ forever or 2) leaves $\S_{[s_0]}$ eventually. Note that $\texttt{v}_\text{stay}(s_0)$ and $\texttt{v}_\text{trans}(s_0)$ record the maximum entropy rates under situations 1) and 2), respectively. Thus \eqref{eq:stayandtransisenough} holds.
\end{proof}

By combining Propositions~\ref{prop:transientoptimal} and~\ref{prop:MECmaximumisstayandtrans}, 
we have the following result immediately for the fused policy $\hat{\mu}_{k+1}$.
\begin{mypro} \label{prop:optimalpro}
Suppose that $\hat{\mu}_k$ is an optimal solution to Problem~1 for instant sub-MDP $\hat{\M}_{k}=(\hat{R}_{k},\hat{A}_{k})$. 
Then policy $\hat{\mu}_{k+1}$ defined in Equation~\eqref{eq:fuse} is an optimal solution to Problem~1 for instant sub-MDP $\hat{\M}_{k+1}=(\hat{R}_{k+1},\hat{A}_{k+1})$. 
\end{mypro}
Finally, we establish the   correctness of Algorithm~\ref{alg:gmdpsolution}.
\begin{mythm} \label{thm:solution}
Given MDP $\mathcal{M}=(S,s_0,A,P,\mathcal{AP},\ell,Acc)$, the output of Algorithm~\ref{alg:gmdpsolution} is a solution of Problem~\ref{problem:maxicons} for $\M$.
\end{mythm}
\begin{proof}
Note that we assume that initial from any $s \in S$, the system can satisfy LTL task w.p.1. 
Thus for any $s \in L_{0}$, MEC $(\S_{[s]},\A_{[s]})$ contains some AMECs. 
By Theorem~\ref{thm:alg1isright}, we know that $\hat{\mu}_0$ in line 4 of Algorithm~\ref{alg:gmdpsolution} is a solution of Problem~\ref{problem:maxicons} w.r.t. $\hat{\M}_{0}$. 
By Proposition~\ref{prop:optimalpro}, we know that at each step $k$ we find solution of Problem~\ref{problem:maxicons} w.r.t. $\hat{\M}_{k+1}$. Since $\hat{\M}_{\texttt{level}(\M)+1}=\M$, we know the return policy solves Problem~\ref{problem:maxicons} for $\M$. 
\end{proof}

\begin{remark}
Note that, for the sake of simplicity, 
in Equation~\eqref{eq:trans-enf}, we construct the transient-enforcing policy $\hat{\mu}_{k+1}'$ as a deterministic policy. This is because we are concerned with maximizing the \emph{entropy rate}, and transient behaviors will not affect this result. If one wants to further maximize the unpredictability of the transient behaviors, then one can adopt the approach in~\cite{savas2019entropy} for the transient part.
\end{remark}

\section{Case Studies of Robot Task Planning}\label{sec:7}
In this section, 
we present two case studies of robot task planning to illustrate the proposed method. 
All computations are performed on a desktop with 16 GB RAM. 
Specifically,  we use the \textsf{splitting conic solver} (SCS)~\cite{o2016conic} in \textsf{CVXPY}~\cite{diamond2016cvxpy} to solve convex optimization problems. 
Also, we use the tool in~\cite{klein2007ltl2dstar} to transform the LTL task to DRA. 

\subsection{Case Study 1}

\begin{figure}
    \centering
    \begin{tikzpicture}

   	\draw[fill=blue, draw = white] (5.4,0.45) -- (5.4,0.75) -- (5.7, 0.75) -- (5.7, 0.45);
   	
   	\draw[fill=blue, draw = white] (6.3,1.95) -- (6.3,2.25) -- (6.6, 2.25) -- (6.6, 1.95);
   	
   	\draw[fill=blue, draw = white] (3,-0.75) -- (3,-1.05) -- (3.3, -1.05) -- (3.3, -0.75);

	\foreach \x in {0,0.3,...,2.4}
	\draw (\x, 1.05)--(\x, -1.05);
	
	\foreach \z in {2.4,2.7,...,5.1}
	\draw (\z,0.15)--(\z,2.55);
	
	\foreach \z in {2.4,2.7,...,5.1}
	\draw (\z,-0.15)--(\z,-2.55);
	
	\foreach \z in {5.1,5.4,...,7.5}
	\draw (\z,0.15)--(\z,2.55);
	
	\foreach \z in {5.1,5.4,...,7.5}
	\draw (\z,-0.15)--(\z,-2.55);
	
	\foreach \y in {-1.05,-0.75,...,1.35}
	\draw[densely dotted] (0, \y)--(2.1, \y);
	
	\foreach \y in {0.15,0.45,...,2.85}
	\draw[densely dotted] (2.4, \y)--(4.8, \y);
	
	\foreach \y in {-0.15,-0.45,...,-2.85}
	\draw[densely dotted] (2.4, \y)--(4.8, \y);
	
	\foreach \y in {-0.15,-0.45,...,-2.85}
	\draw[densely dotted] (5.1, \y)--(7.5, \y);
	
	\foreach \y in {0.15,0.45,...,2.85}
	\draw[densely dotted] (5.1, \y)--(7.5, \y);
	
	\draw[densely dotted] (2.1,1.05) -- (2.4,1.05);
	
    \draw[densely dotted] (2.1,0.75) -- (2.4,0.75);
    
	\draw[densely dotted] (4.8,1.05) -- (5.1,1.05);
	
    \draw[densely dotted] (4.8,0.75) -- (5.1,0.75);
    
	\draw[densely dotted] (2.1,-1.05) -- (2.4,-1.05);
	
    \draw[densely dotted] (2.1,-0.75) -- (2.4,-0.75);
    
	\draw[densely dotted] (4.8,-1.05) -- (5.1,-1.05);
	
    \draw[densely dotted] (4.8,-0.75) -- (5.1,-0.75);
    
    \draw[densely dotted] (4.8,-1.95) -- (5.1,-1.95);
	
    \draw[densely dotted] (4.8,-1.65) -- (5.1,-1.65);
    
    \draw (6,0.15) -- (6,-0.15); 
    
    \draw (6.3,0.15) -- (6.3,-0.15);
    
    \draw [line width =1.5pt] (0, 1.05)--(0, -1.05);
    \draw [line width =1.5pt] (2.1, 0.75)--(2.1, -0.75);
    
    \draw [line width =1.5pt] (2.4, 0.15)--(2.4, 0.75);
    \draw [line width =1.5pt] (2.4, 1.05)--(2.4, 2.55);
    
    \draw [line width =1.5pt] (2.4, -0.15)--(2.4, -0.75);
    \draw [line width =1.5pt] (2.4, -1.05)--(2.4, -2.55);
    
    \draw [line width =1.5pt] (4.8, 0.15)--(4.8, 0.75);
    \draw [line width =1.5pt] (4.8, 1.05)--(4.8, 2.55);
    
    \draw [line width =1.5pt] (4.8, -0.15)--(4.8, -0.75);
    \draw [line width =1.5pt] (4.8, -1.05)--(4.8, -1.65);
    \draw [line width =1.5pt] (4.8, -1.95)--(4.8, -2.55);
    
    \draw [line width =1.5pt] (5.1, 0.15)--(5.1, 0.75);
    \draw [line width =1.5pt] (5.1, 1.05)--(5.1, 2.55);
    
    \draw [line width =1.5pt] (5.1, -0.15)--(5.1, -0.75);
    \draw [line width =1.5pt] (5.1, -1.05)--(5.1, -1.65);
    \draw [line width =1.5pt] (5.1, -1.95)--(5.1, -2.55);
    
    \draw [line width =1.5pt] (7.5, 0.15)--(7.5, 2.55);
    \draw [line width =1.5pt] (7.5, -0.15)--(7.5, -2.55);
    
    \draw [line width =1.5pt] (0, 1.05)--(2.4, 1.05);
    \draw [line width =1.5pt] (0, -1.05)--(2.4, -1.05);
    
    \draw [line width =1.5pt] (2.1, -0.75)--(2.4, -0.75);
    \draw [line width =1.5pt] (2.1, 0.75)--(2.4, 0.75);
    
    \draw [line width =1.5pt] (2.4, 0.15)--(4.8, 0.15);
    
    \draw [line width =1.5pt] (5.1, 0.15)--(6, 0.15);
    \draw [line width =1.5pt] (6.3, 0.15)--(7.5, 0.15);
    
    \draw [line width =1.5pt] (2.4, -0.15)--(4.8, -0.15);
    
    \draw [line width =1.5pt] (5.1, -0.15)--(6, -0.15);
    \draw [line width =1.5pt] (6.3, -0.15)--(7.5, -0.15);
    
    \draw [line width =1.5pt] (2.4, 2.55)--(4.8, 2.55);
    \draw [line width =1.5pt] (5.1, 2.55)--(7.5, 2.55);
    
    \draw [line width =1.5pt] (2.4, -2.55)--(4.8, -2.55);
    \draw [line width =1.5pt] (5.1, -2.55)--(7.5, -2.55);
    
    \draw [line width =1.5pt] (4.8, 0.75)--(5.1, 0.75);
    \draw [line width =1.5pt] (4.8, 1.05)--(5.1, 1.05);    
    \draw [line width =1.5pt] (4.8, -0.75)--(5.1, -0.75);
    \draw [line width =1.5pt] (4.8, -1.05)--(5.1, -1.05);
    
    \draw [line width =1.5pt] (4.8, -1.65)--(5.1, -1.65);
    \draw [line width =1.5pt] (4.8, -1.95)--(5.1, -1.95);
    
    \draw [line width =1.5pt] (6, -0.15)--(6, 0.15);
    \draw [line width =1.5pt] (6.3, -0.15)--(6.3, 0.15);

    \draw[-{Latex}] (-0.35,0) -- (-0.05,0);
    
    \node (1) at (1.05,0){$1$};
    \node (2) at (3.75,1.2){$2$};
    \node (3) at (3.75,-1.5){$3$};
    \node (4) at (6.45,1.2){$4$};
    \node (5) at (6.45,-1.5){$5$};
    
    	\begin{scope}[red, line width=0.3mm, arrows = {-Latex[scale=0.6]}]
    	\draw (2.3,0.9) -- (2.6,0.9);
    	\draw (2.3,-0.9) -- (2.6,-0.9);
    	\draw (4.9,-0.9) -- (4.6,-0.9);
    	\draw (5,-1.8) -- (5.3,-1.8);
    	\draw (5,0.9) -- (5.3,0.9);
    	\draw (6.15,0.05) -- (6.15,0.35);
    	\end{scope}
\end{tikzpicture}
    \caption{Workspace of the robot.}
    \label{fig:case1}
\end{figure}
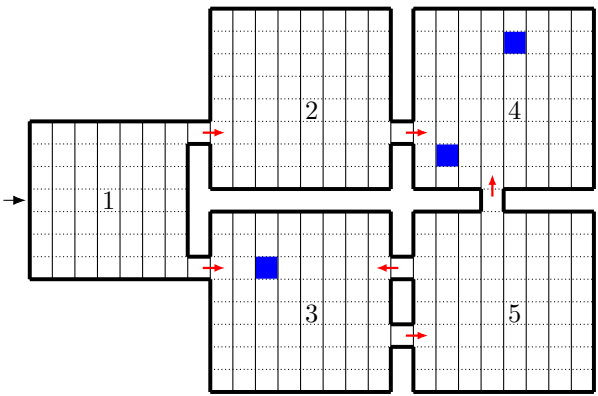

\textbf{System Model: }  
We consider a robot moving in a workspace shown in  Figure~\ref{fig:case1}.
The entire workspace consists of five regions, 
where Region~1 consists of $7\times 7$ grids and each of Regions~$2$-$5$ consists of $8\times 8$ grids. 
The initial location of  the robot is pointed by the black arrow.
The five regions are connected by some one-way path grids whose feasible directions are depicted in the figure, e.g., robot can reach regions 2 and 3 from region 1 but cannot reach region 1 from regions 2 or 3. 
The mobility of the robot is as follows. 
Inside of each region, the robot has five actions, left/right/up/down/stay. 
By choosing each action, the robot will move to the target grid w.p.1. 
Furthermore, if the robot chooses an action but the target grid is a wall (the boundary of the region), then it will stay in the current grid. 
Between two regions, the robot can only move through the one-way path grids following the given directions. 
Therefore, the mobility of the robot can be modeled as an MDP $\mathcal{M}$  (in fact, deterministic) with $310$ states and $1379$ edges. 
Clearly, there are four MECs in $\mathcal{M}$, where Regions~$3$ and $5$ belong to the same MEC.

\textbf{LTL Task: }
We assume that the robot needs to visit some specific grids, which are marked by blue color with label $b$, infinitely often in order to communicate with the center station.  
Then the  task is  simply
$\varphi = \square \lozenge b$.

\begin{figure} 
  \centering
  \includegraphics[width=8cm]{ 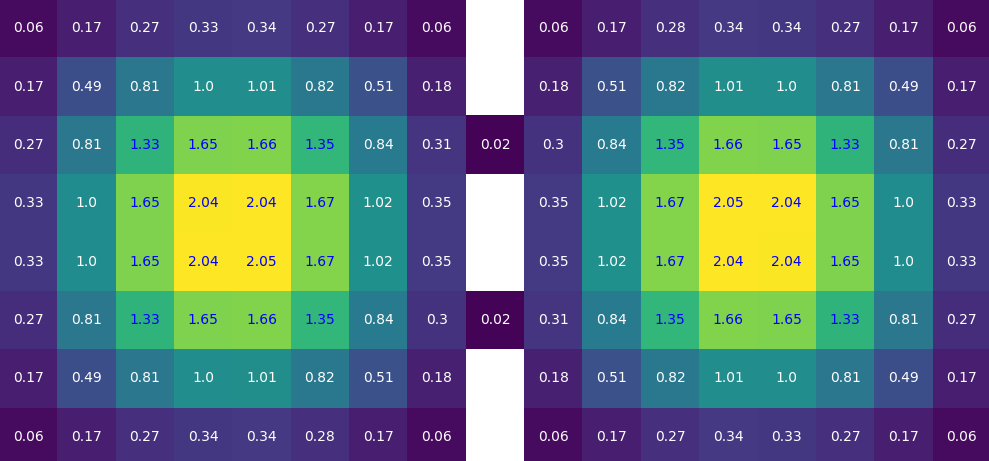}
  \caption{Limit distribution (multiplied by $100$) of the optimal policy $\mathcal{M}^{\mu}$.}
 \label{fig:limitdis1}
\end{figure}

\begin{figure} 
  \centering
  \includegraphics[width=8cm]{ 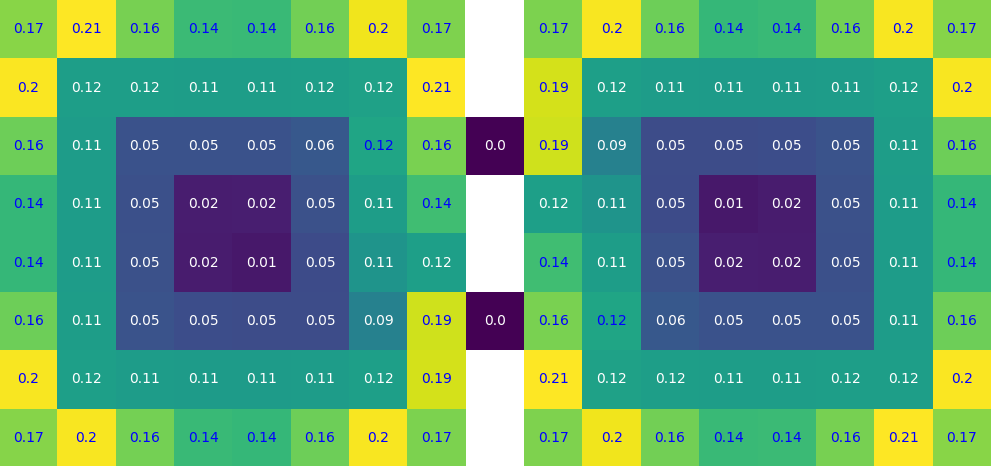}
  \caption{Largest difference of the probabilities of picking two different actions at each  state.}
 \label{fig:diff}
\end{figure}

\textbf{Results:}
Note that, the product MDP for this task is isomorphic to the original MDP, 
and there are two AMECs in $\mathcal{M}$:  Region $4$, and the union of Regions $3$ and $5$. 
We denote by $\mu$ the solution to Problem~\ref{problem:maxicons}. 
Clearly, the robot will choose to eventually stay in Regions~$3$ and $5$ rather than stay in Region~4 only. This is because staying a larger region will increase its entropy rate. 
The limit distribution (multiplied by 100) of each state under $\mu$ is shown in Figure~\ref{fig:limitdis1}. Note that, it suffices to show  Regions~$3$ and $5$ as the limit distribution of states in other regions are zero.
One may think that the maximum entropy rate policy is uniformly randomized. 
However, it is not the case. 
Specifically, for each state in Regions 3 and 5, 
we compute the difference between highest probability of an action and the lowest probability of an action in the optimal policy; the difference values are shown in Figure~\ref{fig:diff}. 
Clearly, only when the robot is at the center of the Region, it will follow a purely randomized strategy. 
For the remaining states, the optimal policy  is not uniformly randomized if it wants to maximize the entropy rate.


\textbf{Comparison:}
In the context of information-theoretical foundation of security, a useful measure for quantifying the unpredictability of an agent is the weight of the Huffman tree of the distribution; see, e.g., \cite{paruchuri2006security}.  
This value has the following physical meaning. Suppose that  there is an observer knowing the structure of the MDP and the policy of the agent. 
In each state, it runs yes-no probes to know the successor state of the agent. If agent moves to state that probe predicts, the probe will return yes; otherwise it returns no. Specifically, let $\Upsilon_s$ be the weight of the Huffman tree for the next-step transition distribution at state $s$. 
Then this value is actually  the running time of yes-no probes in state $s$. 
Therefore, the \emph{average number of observations} needed to determine the path of agent under policy $\mu$ can be characterized by 
$O_{a}^{\mu}=\sum_{s} \pi^\mu(s) \Upsilon_{s}$ where $\pi^\mu$ is limit distribution of MC $\M^\mu$. 
If we adopt the algorithm in \cite{savas2019entropy} to synthesize a policy $\mu_1$ that finishes the task w.p.1 and maximizes the total entropy of MDP, then it holds that $O_{a}^{\mu_1}=0$ since $\mu_1$ will choose a deterministic action in the steady state. 
However, for our algorithm, we have  $O_{a}^{\mu}=2.56$. Therefore, method proposed makes the limit behavior of agent more unpredictable.

\subsection{Case Study 2}

\begin{figure}  
\subfigure[Workspace of the robot] 
	{\label{fig:workspace2}
            \begin{minipage}[b]{0.46\linewidth}
               	\centering
\includegraphics[height=4.8cm]{ 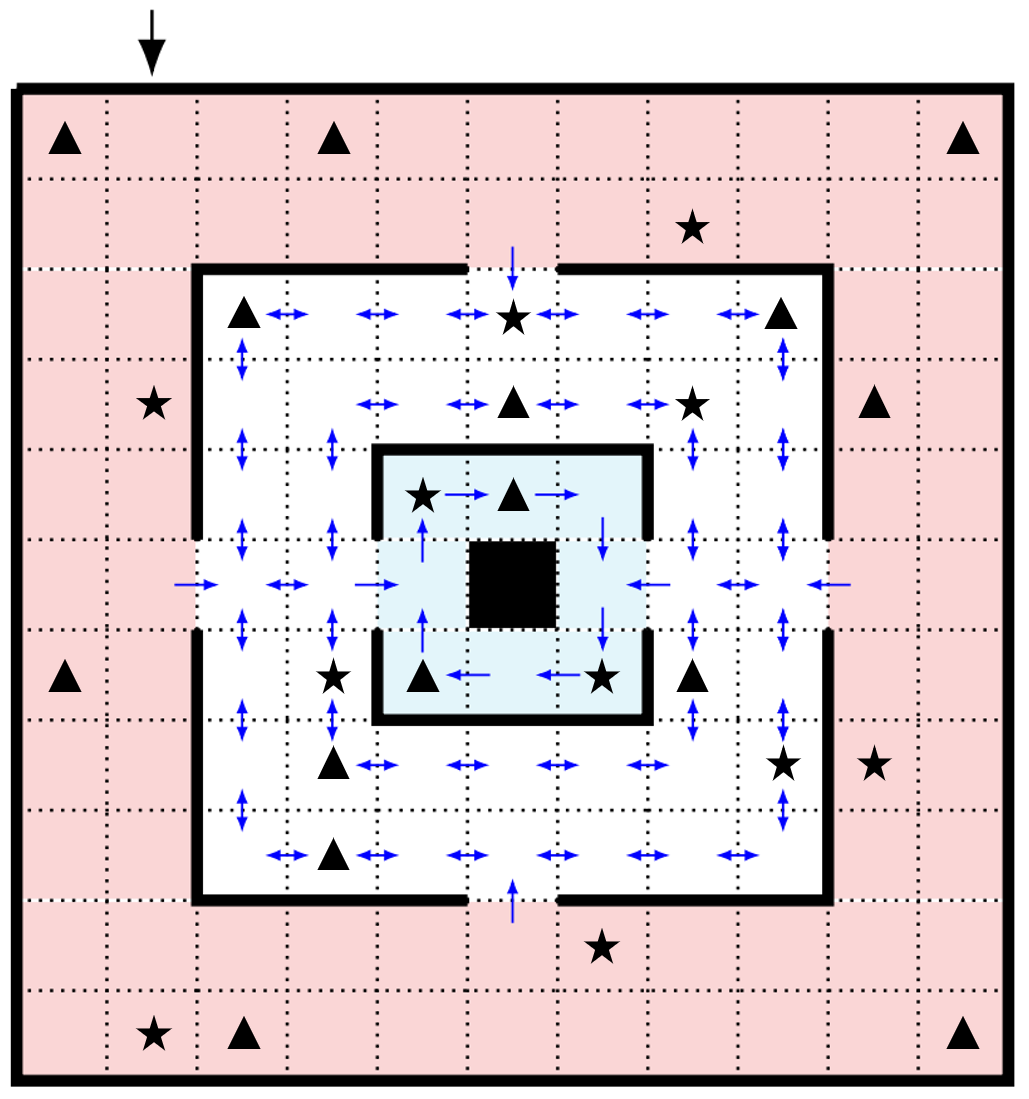}
           \end{minipage}
	}
	\subfigure[Limit distributions (multiplied by $100$) of  three AMECs. ] 
	{\label{fig:solution3AMEC}
 \begin{minipage}[b]{0.46\linewidth}
	\centering
 \includegraphics[height=4.48cm]{ 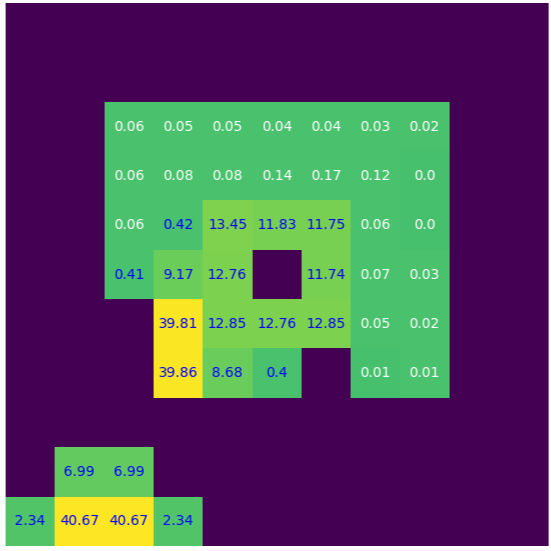}
    \end{minipage}
	}
	
    \caption{Workspace and optimal policy for Case Study 2.}
   \label{fig:case2}
\end{figure}

\textbf{System Model:} 
We consider a data collecting and uploading scenario where the robot works in an $11\times 11$ grids as shown in Figure~\ref{fig:workspace2}.
Specifically, the overall workspace consists of three regions segmented by walls, denoted by solid black lines in the figure. 
Region 1 is the light blue area in the center,
Region 2 is the white area in the middle, and
Region 3 is the light red area on the outside.
The \emph{initial} location of the robot is the upper left grid, indicated by the black arrow.

We assume that the robot will have more mobility constraints in the internal regions. Specifically, it can move to adjacent grids freely in Region 3. However, in Regions 1 and 2, the robot can only choose to stay in its current grid or move in the directions indicated by the blue arrows. 
To move from Region 3 to Region 2, the robot can only use one of the four one-way openings in the wall. Similarly, to move from Region 2 to Region 1, the robot can only use one of the two one-way openings in the wall. Therefore, eventually, the robot will stay within one region forever and cannot travel between regions interchangeably.

To connect this problem with security considerations, we assume there exists an observer who knows the structure of the MDP and the agent's policy. At each state, the observer conducts yes-no probes to determine the agent's successor state. The number of observations required by the observer corresponds to the number of yes-no probes performed in that state. 
\emph{The more observations needed, the greater the computational resources required to obtain the robot's state information, and the harder it becomes for the observer to cause harm.}  
The observer employs the Huffman procedure~\cite{huffman1952method} to minimize the expected number of probes. 
Let $\Upsilon_s$ denote the weight of the Huffman tree for the next-step transition distribution at state $s$. Then, the \emph{average expected number of observations} required to determine the agent's path under policy $\mu$ is given by   
$ O_{a}^{\mu} = \sum_{s} \pi^\mu(s) \Upsilon_{s}$,   
where $\pi^\mu$ is the stationary distribution of the Markov chain $\mathcal{M}^\mu$. For further details on this observer model, we refer the reader to~\cite{paruchuri2006security}.

\textbf{LTL Task:} 
We assume that in the workspace, there are two types of data stations, denoted by $\bigstar$ and $\blacktriangle$, respectively, where data is collected and uploaded. Specifically, information $b$ can only be collected at station $\bigstar$ and uploaded at station $\blacktriangle$. Conversely, data $r$ can only be collected at station $\blacktriangle$ and uploaded at station $\bigstar$.
For each region $i=1,2,3$, there exists a time bound $t_i$ such that, once  data is collected within this region, it must be uploaded within this time interval. 
We denote by $b_i$ and $r_i$ as the atomic propositions such that the grid is a $\bigstar$ station  and a $\blacktriangle$ station   in Region $i$, respectively. 
This communication constraint can be described by the following LTL formula
\begin{equation}
\varphi_i(t_i) = \square ( b_i \to  \bigvee_{j=0}^{t_i} \bigcirc^{j} r_i  ) \wedge \square ( r_i \to  \bigvee_{j=0}^{t_i} \bigcirc^{j} b_i  ),   \nonumber
\end{equation}
where $ \bigcirc^{j}$ means $\underbrace{ \bigcirc\cdots \bigcirc }_{j\text{ times}}$. 
We consider the following time bounds for each region: 
$t_1=8$, $t_2=5$ and $t_3=3$. 
The overall objective is to collect and upload each type of data infinitely often while satisfying the time bound constraints, i.e., 
\begin{equation} \label{eq:exm-ltl}
    \phi =\bigvee_{i=1,2,3} (\square\lozenge b_i\wedge \square\lozenge r_i \wedge \varphi_i(t_i) ).
\end{equation}

Clearly, the robot will choose to stay within one region forever. 
There are also three  AMECs in product MDP corresponding to each grid region. 
We compute the optimal staying policy for each AMEC, 
and the maximum entropy rates for AMECs corresponding to  Regions 1, 2 and 3 are 
$0.68$, $0.86$ and $0.81$, respectively. 
Clearly,  the optimal overall policy  of the robot is to move to Region~2 and stay there forever. 
We show in Figure~\ref{fig:solution3AMEC} the limit distributions (projected from product states to grid states, and multiplied by 100) of each AMEC under each optimal staying policy. Note that, for Region~3, to fulfill the LTL task, the robot can only move around the lower left corner, 
since $t_3=3$ and stations $\bigstar$ and $\blacktriangle$ are too far away from each other in other places.

To better demonstrate how the LTL task affects the optimal policy, we further consider the LTL task $\square\lozenge b_i\wedge \square\lozenge r_i \wedge \varphi_i(t_i) $ for different regions  $i=1,2,3$ and different time bounds $t_i$. The average number of observations and maximum entropy rate of the policy achieving the LTL task for each pair of parameters are shown in Table~\ref{tab:outcomefordiffpara}, where TB represents the time bound between collection and upload, ANO represents the average number of observations and MER represents the maximum entropy rate. 
Note that if the robot adopts a deterministic policy, then the average number of observation is $0$ and the observer can get the state information of robot with no cost.
Clearly, for each region, as the time bound increases, the maximum entropy rate and average number of observations also increase because the robot has more flexibility to be unpredictable. Therefore, if one considers the LTL task in the form of Equation~\eqref{eq:exm-ltl}, the region in which the robot eventually chooses to stay depends on the time bound value $t_i$ for each region.
\begin{table}[tpb]
      \caption{MER and ANO values for different parameters. R$i$ is Region $i$. TB is time bound. First line: MER. Second line: ANO.}
  \label{tab:outcomefordiffpara}
  \centering
  \begin{threeparttable}[b]
     \begin{tabular}{cccccccccc}
      \toprule
     TB  & $1$ & $2$ & $3$ & $4$ & $5$  & $6$ & $7$ & $8$ &$\infty$  \\ 
      \midrule
     \multirow{2}*{R$1$} & - & - & 0.32 & 0.52 & 0.60 & 0.64 & 0.67 &0.68 & 0.69  \\
     & - & - & 0.50 & 0.80 & 0.91 & 0.95 & 0.97 &0.99 & 1.00  \\
     \midrule
     \multirow{2}*{R$2$} & 0 & 0.48 & 0.69 & 0.80 & 0.86 & 0.90 & 0.92 &0.94  &1.18 \\
     & 0 & 0.72 & 1.07 & 1.23 & 1.32 & 1.37 & 1.421 &1.423 &1.75 \\
     \midrule
      \multirow{2}*{R$3$} & 0 & 0.48 & 0.81 & 0.95 & 1.04 & 1.06 & 1.07 &1.11 &1.40 \\
      & 0 & 0.72 & 1.20 & 1.43 & 1.59 & 1.67 & 1.76 &1.91 &2.23 \\
      \bottomrule
    \end{tabular}

  \end{threeparttable}
\end{table}

\section{Conclusion}\label{sec:8}
In this paper, we solved a new entropy rate maximization problem for MDPs under the requirement that a given linear temporal logic task needs to be achieved  with probability one. We first solved this problem for special case of communicating MDPs by an efficient convex optimization problem. 
For general MDPs, we showed that this problem can be effectively
solved by decomposing it as a finite set of sub-problems by proposed state level classification method. Our results extended  existing results in entropy rate maximization by taking temporal logic constraints into account. 
We demonstrated the proposed algorithm by two case studies of robot task planning. 
In the future, we would like to further investigate how to solve this problem under the partial observation setting. 

\appendix \label{appendix}

Our proof latter needs to leverage the results from MDPs with expected total reward. 
Specifically, let  $\mathcal{R}: S \times A \rightarrow \mathbb{R}$ be a reward function that maps each state-action pair to a real number. Given policy $\mu \in \Pi_{\mathcal{M}}$, the \emph{expected total} reward starting from $s \in S$ is defined by
    \begin{equation}
      v^{\mu}_{\mathcal{M}}(s)=  E_{s}^{\mu} \left[ \sum_{t=0}^{\infty} \mathcal{R}(S_{t},A_{t}) \right].
  \end{equation}
The reward vector for all states is denoted by $v^{\mu}_{\mathcal{M}}$. 
For $s \in S$, we define $v^{\star}_{\mathcal{M}}(s)=\sup_{\mu \in \Pi_{\mathcal{M}}} v^{\mu}_{\mathcal{M}}(s)$. A policy $\mu^{\star} \in \Pi_{\mathcal{M}}$ is optimal if $v^{\star}_{\mathcal{M}}(s)=v^{\mu^{\star}}_{\mathcal{M}}(s)$ for all $s \in S$. We omit subscript $\mathcal{M}$ if it is clear from context.

Note that, for each AMEC $(\hat{\S},\hat{\A}) \in \texttt{AMEC}(\M)$, we can find an MEC $(\S,\A) \in \texttt{MEC}(\M)$ such that $\hat{\S} \subseteq \S$ and $\hat{\A}\subseteq \A$. 
We denote by $\texttt{MEC}_{\varphi}(\M)$ the set of MECs  containing at least one AMEC. 
For a set of MECs $\texttt{M} \subseteq \texttt{MEC}_{\varphi}(\M)$, 
we denote by $\mathcal{S}_\texttt{M} \subseteq S$ the set of all states in $\texttt{M}$.
Also,  for each MEC $(\S,\A) \in \texttt{MEC}_\varphi(\M)$, 
we define 
\[
V^\star(\S,\A)=\max_{(\hat{\S},\hat{\A}) \in \texttt{AMEC}(\S,\A)} \nabla H(\mathcal{M} (\hat{\S},\hat{\A}) )
\] 
as the maximum entropy rate one can achieve among all AMECs in $(\S,\A)$. Let $(\hat{\S},\hat{\A}) \in \texttt{AMEC}(\S,\A)$ be the AMEC achieving $V^\star(\S,\A)$. Define $\mu_{(\S,\A)}\in \Pi^S_{\M(\S,\A)}$ s.t. for $s\in \hat{\S}$, $\mu_{(\S,\A)}(s,a)=\mu_{(\hat{\S},\hat{\A})}^\star(s,a)$ with $\mu_{(\hat{\S},\hat{\A})}^\star$ the maximum entropy rate policy of sub-MDP $(\hat{\S},\hat{\A})$, and for states set $\S \setminus \hat{\S}$, $\mu_{(\S,\A)}$ ensures that $\S \setminus \hat{\S}$ can reach $\hat{\S}$ eventually w.p.1 \cite[Page 480]{puterman}. Then $\mu_{(\S,\A)}$ achieves $V^\star(\S,\A)$ over sub-MDP $(\S,\A)$.

Given MDP $\M$ with state space $S$ and $\hat{S} \subseteq S$, let
\begin{equation}\label{eq:newMDP}
    \M_{\hat{S}}=(\Bar{S},\Bar{A},\Bar{P})
\end{equation}
be an MDP such that 
$\Bar{S}=S \cup \{ b \}$ with $b$ be a new state, 
$\Bar{A}(s)=A(s)$ for $s \in S \setminus \hat{S}$ and 
$\Bar{A}(s)=\{\Bar{a}\}$ for $s \in \hat{S} \cup \{ b\}$ with $\Bar{a}$ be a new action. 
For $s \in S\setminus \hat{S}$, we define $\Bar{P}_{s,a,t}=P_{s,a,t}$ and 
for $s \in \hat{S} \cup \{ b\}$, we define $\Bar{P}_{s,\Bar{a},b}=1$. Given MECs set $\texttt{M} \subseteq \texttt{MEC}_{\varphi}(\M)$, we define a reward function $\mathcal{R}_{\texttt{M}}$ s.t.
\begin{align}\label{eq:reward}
	\mathcal{R}_{\texttt{M}}(s,a) \!= \!
		\left\{\!\!
		\begin{array}{cl}
			V^\star(\S_{[s]},\A_{[s]})  & \text{if }    s \in  \S_{\texttt{M}} \\
         0      & \text{otherwise}  
		\end{array}
		\right.\!\!\!.
\end{align}
 From \cite{puterman} there exists a stationary policy, denoted by $\mu_{\texttt{M}}^\star \in 
 \Pi^S_{\M_{\S_{\texttt{M}}}}$, 
for total expected reward maximization w.r.t. $\M_{\S_\texttt{M}}$ and $\mathcal{R}_{\texttt{M}}$ under any initial distribution. We denote by $\tilde{\S}_{\texttt{M}} \subseteq S$ the set of states which can reach states in $\S_{\texttt{M}}$ w.p.1 under some policy. Since $V^\star(\S,\A) \geq 0$ for any $(\S,\A) \in \texttt{MEC}_\varphi(\M)$, we can assume without loss of generality that states in $\tilde{\S}_{\texttt{M}}$ will reach $\S_{\texttt{M}}$ w.p.1 in MC $\M^{\mu_{\texttt{M}}^\star}_{\S_{\texttt{M}}}$, because it can only get zero reward when reaching MECs not in $\texttt{M}$ under reward $\mathcal{R}_{\texttt{M}}$. For $\texttt{M} \subseteq \texttt{MEC}_\varphi(\M)$, 
we denote by $\mu_{\texttt{M}} \in \Pi^S_\M$ the policy such that $\mu_{\texttt{M}}(s,a) = \mu^\star_{\texttt{M}}(s,a)$ if $s \in S\setminus \S_{\texttt{M}}$ and $\mu_{\texttt{M}}(s,a)=\mu_{(\S_{[s]},\A_{[s]})}(s,a)$ if $s\in \S_{\texttt{M}}$. We define by
\begin{equation}\label{eq:collectionofstationary}
    \texttt{col}=\{ \mu_{\texttt{M}} \in \Pi_{\M}^S \mid \texttt{M} \subseteq \texttt{MEC}_\varphi(\M), s_0 \in \tilde{\S}_{\texttt{M}} \} \subseteq \Pi^\varphi_\M
\end{equation}
the set of policies $\mu_{\texttt{M}}$ such that MC $\M^{\mu_{\texttt{M}}}$ will reach $\S_\texttt{M}$ w.p.1 from initial state $s_0$.

We start by proving Lemma~\ref{prop:com_irr}. 
To this end, we need the following auxiliary result showing that the entropy rate for MC is a linear combination of that for each recurrent class.
\begin{mycla} \label{prop:limitdistri}
Given MDP $\mathcal{M}$ and policy $\mu \in \Pi^{S}_\mathcal{M}$, 
suppose that $\mathcal{M}^{\mu}$ has $K$ recurrent classes $R_{1},\dots,R_{K} \subseteq S$. 
Let $E_k$ be the entropy rate for the MC restricted on $R_k$. 
Then there exists a set of values $\beta(1),\dots, \beta(K)\in [0,1]$ such that $\sum_{i=1}^K \beta(k)=1$ and
\[
\nabla H(\M^\mu) = \sum_{k=1}^K \beta(k) E_k.
\]
\end{mycla}
\begin{proof}
Let $T=S\setminus \bigcup_{k=1}^K R_k$ be the set of transient states. 
Let $Q_0$ and $Q_k$ be the transition matrices from $T$ to $T$ and $R_k$, respectively. 
Let $\pi_0$ be the row vector of the initial distribution.
According to \cite[page 593]{puterman}, we have $\sum_{s \in R_k} \pi(s) L(s)=\beta(k) E_k$ and
\[
\beta(k) =\left( \pi^T_0 (I-Q_{0})^{-1}Q_{k}+\pi^k_0 \right ) \mathbf{e},
\]
where $\mathbf{e}$ denotes the one-vector with suitable dimension, 
$\pi^T_0$ and $\pi^k_0$ are the initial distributions restricted on $T$ and $R_k$, respectively.
The probability of reaching $R_k$ from $s \in T$ is
\begin{equation} \label{eq:otherpropmiddle}
    (I-Q_0)^{-1} Q_k \mathbf{e}.
\end{equation}
Therefore, we have $\sum_{k=1}^K(I-Q_0)^{-1} Q_k \mathbf{e} = \mathbf{e}$, which means that $\sum_{k=1}^K\beta(k)=1$. 
Note that $\mathbf{e}$ in left and right hand side have different dimensions.
This completes the proof.
\end{proof}

\begin{proof}[\bf Proof of Lemma~\ref{prop:com_irr}]
By Claim~\ref{prop:limitdistri},  the entropy rate is a linear combination of each recurrent class entropy rate. If there are several recurrent classes, then they must have the same entropy rate. Since MDP is communicating, we can select one recurrent class and make other states reach it w.p.1 by procedure in \cite[Page 480]{puterman}. This new policy has same entropy rate as original policy. Thus, without loss of generality, we assume that $\M^\mu$  only has one recurrent class $R$.

Now we prove our result by contradiction. 
Assume that MC $\mathcal{M}^{\mu}$ is not irreducible, i.e., state set $S \setminus R$ is not empty. Since $\mathcal{M}$ is communicating, there exist $r \in R$ and $\hat{a} \in A(r)$ such that $P_{r,\hat{a},\hat{t}} > 0 $ for some $\hat{t} \notin R$. 
We define policy $\hat{\mu}$ such that $\hat{\mu} (s,a)=\mu(s,a)$ for $s \neq r$ and $\hat{\mu}(r,\hat{a})=1$. Also, we define 
\[
\mu_{\epsilon}=(1-\epsilon) \mu + \epsilon \hat{\mu}\text{, where } \epsilon \in [0,1).
\]
Define $\mathbf{d} \in \mathbb{R}^{|S|}$ such that $\mathbf{d}(s)=\mathbb{P}^{\hat{\mu}}_{r,s}-\mathbb{P}^{\mu}_{r,s}$.
Since MC $\mathcal{M}^{\mu}$ only contains one recurrent class and some transient states, by (8) of~\cite{schweitzer1968perturbation}, we have
\begin{equation} \label{eq:limitmiddle}
    \pi^{\mu_{\epsilon}}(s)=\pi^{\mu}(s)+\pi^{\mu}(r) \frac{\epsilon \mathbf{u}(s)}{1-\epsilon \mathbf{u}(r)}, 
\end{equation}
where $\mathbf{u} \in \mathbb{R}^{|S|}$ is the vector such that $\mathbf{u} = \mathbf{d}\mathbf{Z}^{\mu}$ and 
\[
\mathbf{Z}^{\mu} = (I-\mathbb{P}^{\mu}+(\mathbb{P}^{\mu})^{\star})^{-1},
\]
where $(\mathbb{P}^{\mu})^{\star}$ is limit matrix of transition matrix $\mathbb{P}^\mu$.  
Then we define $G(\epsilon)= \nabla H(\M^{\mu_{\epsilon}})$ as a function of $\epsilon$.
By~(\ref{def:localentropy}), we have
 \[
 G(\epsilon)= \pi^{\mu_{\epsilon}}(r)L^{\mu_{\epsilon}}(r) +  \sum_{s \in S, s \neq r} \pi^{\mu_{\epsilon}}(s)L^{\mu_{\epsilon}}(s).
 \]
Let $ G_{2}(\epsilon)= \pi^{\mu_{\epsilon}}(r)L^{\mu_{\epsilon}}(r)$ and $G_1(\epsilon)=G(\epsilon)-G_2(\epsilon)$.
  Since $L^{\mu_{\epsilon}}(s)=L^{\mu}(s)$ for $s \in S$ such that $s \neq r$, the  derivative of $G_1(\epsilon)$ is
  \begin{equation} \label{eq:bounded}
       G'_{1}(\epsilon)=\sum_{s \in S, s\neq r} \frac{\pi^{\mu}(r) \mathbf{u}(s)}{(1-\epsilon \mathbf{u}(r))^{2}} L^{\mu}(s).
  \end{equation} 
Therefore, there exists $ \epsilon_1\in [0,1)$ such that for any $\epsilon \in [0, \epsilon_1]$, $ G'_{1}(\epsilon)$ is bounded.

Similarly, for $G_{2}(\epsilon)$, its derivative is 
\[
G'_{2}(\epsilon)=(\pi^{\mu_{\epsilon}})'(r)L^{\mu_{\epsilon}}(r) + \pi^{\mu_{\epsilon}}(r)(L^{\mu_{\epsilon}})'(r). 
\]
Let $G_{4}(\epsilon) = \pi^{\mu_{\epsilon}}(r)(L^{\mu_{\epsilon}})'(r) $ and 
$G_{3}(\epsilon)=G'_{2}(\epsilon) - G_4(\epsilon)$.
Similarly to \eqref{eq:bounded}, we know that, there exists $ \epsilon_2\in [0,1)$ such that for any $\epsilon \in [0, \epsilon_2]$, $ G_{3}(\epsilon)$ is bounded.

Now we focus on $G_{4}(\epsilon)$. We define 
\[
\mathbb{P}_{s}(\epsilon)=\mathbb{P}^{\mu}_{r,s}-\mathbb{P}^{\hat{\mu}}_{r,s}+(\mathbb{P}^{\mu}_{r,s}-\mathbb{P}^{\hat{\mu}}_{r,s})\log (((1-\epsilon)\mathbb{P}^{\mu}_{r,s}+\epsilon \mathbb{P}^{\hat{\mu}}_{r,s} )) 
\]
and obtain $G_4(\epsilon) = \pi^{\mu_{\epsilon}}(r)(\sum_{s \in S} \mathbb{P}_{s}(\epsilon))$.
We define $T=\{s \in S \mid \mathbb{P}^{\hat{\mu}}_{r,s} > 0 \wedge \mathbb{P}^{\mu}_{r,s} = 0\}$ which is the set of states that can reach from $r$ under $\hat{\mu}$ but cannot reach from $r$ under $\mu$ by one step. Note that $\hat{t} \in T$, i.e., $T$ is non-empty. For $s \in S \setminus T$,  $\mathbb{P}_{s}(\epsilon)$ is bounded for sufficiently small $\epsilon$.
 For $s \in T$, we have
\[
\mathbb{P}_{s}(\epsilon)= -\mathbb{P}^{\hat{\mu}}_{r,s}-\mathbb{P}^{\hat{\mu}}_{r,s}\log (\epsilon\mathbb{P}^{\hat{\mu}}_{r,s}).
\]
Therefore, $\mathbb{P}_{s}(\epsilon)\rightarrow +\infty$ as $\epsilon \rightarrow 0$. By \eqref{eq:limitmiddle}, we know that $\lim_{\epsilon \to 0} G_{4}(\epsilon)  = \pi^\mu(r) \lim_{\epsilon \to 0} \mathbb{P}_{s}(\epsilon)=+\infty$.

In summary, we have 
\[
G'(\epsilon)=G'_{1}(\epsilon)+G_3(\epsilon)+G_4(\epsilon).
\]
Note  we have shown that $G'_{1}(\epsilon)$ and $G_3(\epsilon)$ are both bounded and $G_4(\epsilon)\rightarrow +\infty $ as $\epsilon \rightarrow 0$.
Therefore, $G'(\epsilon) \rightarrow +\infty$ as $\epsilon \rightarrow 0$. 
By Mean Value Theorem~\cite[Thm. 5.10]{rudin1964principles} there exists $\epsilon_{0} > 0$  such that $\nabla H(\M^{\mu_{\epsilon_{0}}})=G(\epsilon_{0}) > G(0)=\nabla H(\mathcal{M}^{\mu})$.
However, it contradicts to the fact that $\mu$ is the maximum entropy rate policy.
\end{proof}

Now, we proceed to prove  Proposition~\ref{prop:suffisenouth}. 
To this end, we still need two auxiliary results, listed as Claims  2 and 3.

For any policy $\mu=(\mu_0,\mu_1,\dots) \in \Pi^\varphi_\M$ and integer $n$, 
we define a new truncated policy 
$\mu^n=(\mu^n_{0},\mu^n_{1},\dots) \in \Pi_{\M}$
such that: 
 (i) for $i< n$, we have $\mu^n_{i}=\mu_i$; and 
(ii) for $i\geq n$, we have 
$\mu^n_{i}(s,a)=\mu_{(\S_{[s]},\A_{[s]})}(s,a)$ if $s \in \S_{\texttt{MEC}_\varphi(\M)}$
and 
$\mu^n_{i}(s,a)=\mu_i(s,a)$ if $s \in S\setminus \S_{\texttt{MEC}_\varphi(\M)}$.
The following result shows that it is without loss of generality to consider such truncated policies.  

\begin{mycla}\label{prop:stationarymiddle1}
Let 
$\hat{\Pi}^\varphi_\M=\bigcup_{\mu \in \Pi^\varphi_\M} \{ \mu^i \}_{i=1}^\infty$
be the set of truncated  policies. 
Then we have $\hat{\Pi}^\varphi_\M \subseteq \Pi^\varphi_\M$ and \begin{equation}\label{eq:twoclassequal}
    \sup_{\mu \in \Pi^\varphi_\M} \nabla H(\M^\mu) = \sup_{\mu \in \hat{\Pi}^\varphi_\M} \nabla H(\M^\mu).
\end{equation}
\end{mycla}
\begin{proof}
For any policy $\mu \in \Pi^{\varphi}_{\M}$, 
AEC $(\hat{\S},\hat{\A}) \in \texttt{AEC}(\M)$ and
MEC $(\S,\A) \in \texttt{MEC}_\varphi(\M)$, 
let 
\begin{align}
\textsf{Pr}^\mu(\hat{\S},\hat{\A})=&\textsf{Pr}^{\mu}_{\M}(\{ \rho \in \textsf{Path}^{\mu}_{\M} \mid \textsf{inf}(\rho) = \hat{\S} \}), \nonumber \\
\textsf{Pr}^\mu_M(\S,\A)=&\sum_{(\hat{\S}',\hat{\A}') \in \texttt{AEC}(\S,\A) }\textsf{Pr}^\mu(\hat{\S}',\hat{\A}') \label{eq:stayingprobability}
\end{align}
be the probability of staying forever in AEC $(\hat{\S},\hat{\A})$ 
and 
the probability of staying forever in AECs in MEC $(\S,\A)$, respectively. 
Then entropy rate under $\mu$ can be expressed as 
\begin{equation} \label{eq:anypolicy}
    \nabla H(\M^\mu) = \sum_{(\hat{\S},\hat{\A}) \in \texttt{AEC}(\M)} \textsf{Pr}^\mu(\hat{\S},\hat{\A}) V^\mu(\hat{\S},\hat{\A}),
\end{equation}
where $V^\mu(\hat{\S},\hat{\A})$ is the entropy rate under $\mu$ restricted on AEC $(\hat{\S},\hat{\A})$.
Note that 
we have
$\sum_{(\hat{\S},\hat{\A}) \in \texttt{AEC}(\M)} \textsf{Pr}^\mu(\hat{\S},\hat{\A})=1$ since $\mu \in \Pi^{\varphi}_{\M}$. 
From \eqref{eq:anypolicy}, we further have
\begin{equation}\label{eq:anypolicybound}
    \nabla H(\M^\mu) \leq \sum_{(\S,\A) \in \texttt{MEC}_\varphi(\M)} \textsf{Pr}^\mu_M(\S,\A)V^\star(\S,\A) .
\end{equation}
 By Lemma~\ref{prop:com_irr}, the  MC  induced by  $\mu_{(\S,\A)}$ will eventually reach some AMEC and visit all states in the AMEC infinitely often. 
 From definition of AMEC, under policy $\mu_{(\S,\A)}$ LTL task $\varphi$ can finish w.p.1 over MEC $(\S,\A)$. 
 Thus $\mu^n \in \Pi^\varphi_\M$ for any $n$, i.e., $\hat{\Pi}^\varphi_\M \subseteq \Pi^\varphi_\M$. 
 Since $\textsf{Pr}^{\mu}_M(\S,\A)=\lim_{n\to \infty}\textsf{Pr}^{\mu^n}_M(\S,\A)$, 
 according to \eqref{eq:anypolicybound}, we know that \eqref{eq:twoclassequal} holds. 
\end{proof}
For $\mu^n \in \hat{\Pi}^\varphi_\M$, let $\texttt{M}(\mu^n)$ be the set of MECs in which it stays forever with non-zero probability. Under policy $\mu^n$, it may happen that: (i) $(\S,\A), (\S',\A') \in \texttt{M}(\mu^n)$ and, (ii) it stays forever in $(\S',\A')$ and stays temporarily at $(\S,\A)$ with non-zero probability. We now further prove that situation (ii) can be prevented without loss of generality.
\begin{mycla} \label{cla:norepeat}
    For $\mu^n \in \hat{\Pi}^\varphi_\M$, if (i) and (ii) hold, we can find $\hat{\mu}^n$ $ \in \hat{\Pi}^\varphi_\M$ such that $\nabla H(\M^{\mu^n}) \leq \nabla H(\M^{\hat{\mu}^n})$ and (ii) is false.
\end{mycla}
\begin{proof}
   From definition of $\mu^n$, we know that under policy $\mu^n$, once reaching $(\S,\A) \in \texttt{MEC}_\varphi(\M)$ at time $i$ with $i \geq n$, it will stay in $(\S,\A)$ forever and execute policy $\mu_{(\S,\A)}$ which achieves entropy rate $V^\star(\S,\A)$. From $\eqref{eq:anypolicy}$, it holds that
    \begin{equation}\label{eq:newcal1}
           \nabla H(\M^{\mu^n}) = \sum_{(\S,\A) \in \texttt{MEC}_\varphi(\M)} \textsf{Pr}^{\mu^n}_M(\S,\A) V^\star(\S,\A).
    \end{equation} 
Then we can select $\hat{\mu}^{\hat{n}} \in \hat{\Pi}^\varphi_\M$ such that for path satisfying (ii),
it only stays forever in either (a)  $(\S,\A)$ or (b) $(\S',\A')$ and chooses higher entropy rate one between (a) and (b). Then $\nabla H(\M^{\mu^n}) \leq \nabla H(\M^{\hat{\mu}^{\hat{n}}})$ and (ii) no longer holds for $\hat{\mu}^{\hat{n}}$.
\end{proof}

Now, we proceed to prove Proposition~\ref{prop:suffisenouth} based on the above two claims. 

\begin{proof}[\bf Proof of Proposition~\ref{prop:suffisenouth}]
For $\mu^n \in \hat{\Pi}^\varphi_\M$, we can repeatedly use Claim~\ref{cla:norepeat} and get a policy $\Bar{\mu} \in \hat{\Pi}^\varphi_\M$ such that for each $(\S,\A) \in \texttt{M}(\Bar{\mu})$, if from initial state $s_0$, it will stay forever in $(\S,\A)$ and stay temporarily at $(\S',\A') \in \texttt{MEC}_\varphi(\M)$ with non-zero probability, then $(\S',\A') \notin \texttt{M}(\Bar{\mu})$. Note that $\texttt{M}(\Bar{\mu})$ is set of MECs it will stay in forever with non-zero probability under $\Bar{\mu}$. Let $\Bar{\mu}' \in \Pi_{\M_{\S_{\texttt{M}(\Bar{\mu})}}}$ be a policy such that $\Bar{\mu}'(s,a)=\Bar{\mu}(s,a)$ for $S\setminus \S_{\texttt{M}(\Bar{\mu})}$ and $\Bar{\mu}'(s,\Bar{a})=1$ for $s \in \S_{\texttt{M}(\Bar{\mu})} \cup \{ b \}$. Let reaching probability of $(\S,\A) \in \texttt{MEC}_\varphi(\M)$ under $\Bar{\mu}'$ be
\begin{equation} \label{eq:reachingpro}
    \textsf{Pr}_R^{\Bar{\mu}'}(\S,\A)=\textsf{Pr}^{\Bar{\mu}'}(\{ \rho \in \textsf{Path}^{\Bar{\mu}'} \mid \exists s \in \S, s \text{ is in } \rho   \}).
\end{equation}
Then it holds that
\begin{equation} \label{eq:newcal2}
     v^{\Bar{\mu}'}(s_0) = \sum_{(\S,\A)\in \texttt{M}(\Bar{\mu})}\textsf{Pr}_R^{\Bar{\mu}'}(\S,\A) V^\star(\S,\A) 
\end{equation}
where $v^{\Bar{\mu}'}$ is total expected reward vector w.r.t. MDP $\M_{\S_{\texttt{M}(\Bar{\mu})}}$ in \eqref{eq:newMDP} and reward $\mathcal{R}_{\texttt{M}(\Bar{\mu})}$ in \eqref{eq:reward}. 
For $(\S,\A) \in \texttt{M}(\Bar{\mu})$, if \eqref{eq:stayingprobability} and \eqref{eq:reachingpro} satisfy $\textsf{Pr}^{\Bar{\mu}}_M(\S,\A)\neq \textsf{Pr}_R^{\Bar{\mu}'}(\S,\A)$, since $\Bar{\mu}$ and $\Bar{\mu}'$ are same over $S \setminus \S_{\texttt{M}(\Bar{\mu})}$, it means that there exists another $(\S',\A') \in \texttt{M}(\Bar{\mu})$ such that it will stay forever in $(\S,\A)$ and stay temporarily at $(\S',\A')$
with non-zero probability. This violates property of $\Bar{\mu}$. Thus $\textsf{Pr}^{\Bar{\mu}}_M(\S,\A)= \textsf{Pr}_R^{\Bar{\mu}'}(\S,\A)$ for $(\S,\A) \in \texttt{M}(\Bar{\mu})$. 
Then combining \eqref{eq:newcal1} and \eqref{eq:newcal2}, it holds that $\nabla H(\M^{\Bar{\mu}})=v^{\Bar{\mu}'}(s_0)$. Therefore, we have
\begin{equation} \label{eq:lastmiddle1}
    \nabla H(\M^{\mu^n}) \leq \nabla H(\M^{\Bar{\mu}}) \leq \nabla H(\M^{\mu_{\texttt{M}(\Bar{\mu})}}) \leq \max_{ \mu_{\texttt{M}} \in \texttt{col}} \nabla H(\M^{\mu_{\texttt{M}}})
\end{equation}
where $\texttt{col}$ and $\mu_{\texttt{M}(\Bar{\mu})}$ are defined in Equation~\eqref{eq:collectionofstationary}. Since $\mu^n$ is selected arbitrary, we know that
\begin{equation} \label{eq:last stepmddle2}
       \sup_{\mu \in \hat{\Pi}^\varphi_\M} \nabla H(\M^\mu)  \leq \max_{ \mu \in \texttt{col}} \nabla H(\M^\mu).
\end{equation}
Furthermore, it is easy to have $\texttt{col} \subseteq \Pi^{\varphi}_\M$. Thus
\begin{equation} \label{eq:last stepmiddle1}
    \max_{ \mu \in \texttt{col}} \nabla H(\M^\mu) \leq \sup_{\mu \in \Pi^\varphi_\M} \nabla H(\M^\mu).
\end{equation}
Finally, combining  \eqref{eq:twoclassequal},  \eqref{eq:last stepmddle2}, \eqref{eq:last stepmiddle1}, we know that one of stationary policies in $\texttt{col}$ is a solution of Problem~\ref{problem:maxicons} w.r.t. $\M$.
This completes the proof.
\end{proof}

Now we prove the Proposition~\ref{pro:optimalCMDP}. We intuitively state its main idea below. First, each $\mu \in \Pi^{S}_\mathcal{M}$ corresponds to a feasible solution of the program. Thus the maximum entropy rate value is not larger than optimal objective value of program. Then for each feasible solution, its objective value is equal to the entropy rate value of some MC under specific initial distribution. Since this entropy rate value is always smaller than specific unichain MC whose entropy rate value is independent with the initial distribution, we know that optimal objective value of program is not larger than the maximum entropy rate value. Combining these, Equation~(\ref{eq:opt-sta}) simply decodes a policy that achieves the best stationary distribution of the desired MC.
\begin{proof}[\bf Proof of Proposition~\ref{pro:optimalCMDP}]
By Proposition~\ref{prop:suffisenouth} we only need to prove that $\mu^\star$ achieves maximum entropy rate over stationary policies. 
Let  $\mu \in \Pi^{S}_\mathcal{M}$ be an arbitrary policy and let $\pi^\mu$ be the limit distribution under $\mu$. 
We define 
$\gamma(s,a)=\pi^\mu(s)\mu(s,a)$, $\lambda(s)=\pi^\mu(s)$ and 
$q(s,t)=\pi^\mu(s)\mathbb{P}^{\mu}_{s,t}$. Clearly, these variables satisfy constraints~(\ref{opt1:con1}), (\ref{opt1:con2}), (\ref{opt1:con4}) and (\ref{opt1:con5}). Since $\pi^\mu \mathbb{P}^{\mu}=\pi^\mu$, we know that constraint~(\ref{opt1:con3}) is also satisfied. Hence, $\{\gamma(s,a)\}_{s\in S, a\in A(s)}$ is a feasible solution to program \eqref{opt1:obj}-\eqref{opt1:con5}. Moreover, we have
\begin{equation}
    \begin{aligned}
         \nabla H(\M^\mu) 
        =& \sum_{s \in S} \pi^\mu(s)L^{\mu}(s)  
        =  \sum_{s \in S} \sum_{t \in S} -\pi^\mu(s)\mathbb{P}^{\mu}_{s,t}\log (\mathbb{P}^{\mu}_{s,t})\\ 
        = & \sum_{s \in S} \sum_{t \in S} -q(s,t)\log \left(\frac{q(s,t)}{\lambda(s)}\right), \nonumber
    \end{aligned}
\end{equation}
i.e., the entropy rate of MC $\mathcal{M}^{\mu}$ is equal to the value of objective function~(\ref{opt1:obj}).
Since no restriction on initial distribution is made, the optimal value of~(\ref{opt1:obj})-(\ref{opt1:con5}), denoted by $V^{\star}$, satisfies that for any initial distribution, $V^{\star} \geq \nabla H(\mathcal{M})$.

On the  other hand, for any $\gamma(s,a)$  satisfying constraints~(\ref{opt1:con1})-(\ref{opt1:con5}), we construct a policy $\mu$ as follows: 
    \begin{equation} 
\mu(s,a)=\frac{\gamma(s,a)}{\sum_{a \in A(s)} \gamma(s,a)} \text{ for } s \in Rc,\label{cpolicy1}  
    \end{equation}
where $Rc=\{ s \in S \mid \sum_{a \in A(s)} \gamma(s,a) > 0 \}$  and $\mu(s,a)$ is assigned arbitrarily  for $s \in S \setminus Rc$. 
By Proposition 9.3.2 of~\cite{puterman}, we know that $Rc$ consists of $K$ recurrent classes in MC $\M^\mu$. 
For each $k=1,\dots ,K$, we denote by $\mathbb{P}^{k}$ the submatrix of $\mathbb{P}^{\mu}$ restricted on recurrent class $R_{k}$. 
For any states $s,t \in R_{k}$, we have
\begin{equation}\label{eq:thm1middle1}
    \mathbb{P}^{k}_{s,t} = \sum_{a \in A(s)} \frac{\gamma(s,a)}{\sum_{a'\in A(s)} \gamma(s,a')} P(t\mid s,a) = \frac{q(s,t)}{\lambda(s)}.
\end{equation}
Then the limit distribution of $R_{k}$ is $(\lambda(s)/\zeta_k)_{s \in R_{k}}$, where $\zeta_{k}=\sum_{s \in R_{k}} \lambda(s)$, since for $t \in R_k$,
\[
\lambda(t) = \sum_{s \in R_k} q (s,t)=\sum_{s\in R_{k}} \lambda(s) \frac{q(s,t)}{\lambda(s)}=\sum_{s \in R_k} \lambda(s)\mathbb{P}^{k}_{s,t}.
\]
The first equality holds since \eqref{opt1:con3} and $q(s,t)=0$ for $s \in S \setminus R_k$. 
The third equality comes from \eqref{eq:thm1middle1}. 
Then the objective function~(\ref{opt1:obj}) with solution $\gamma(s,a)$ can be written as
\begin{equation} \label{upf:1}
    \sum_{k=1}^{K} \sum_{s \in R_{k}} \sum_{t \in R_{k}} -\lambda(s)\mathbb{P}^{\mu}_{s,t}\log (\mathbb{P}^{\mu}_{s,t}).
\end{equation}
The value in (\ref{upf:1}) is also equal to the entropy rate of MC $\mathcal{M}^{\mu}$ with initial distribution $\pi'_{0}$ satisfying $\sum_{s \in R_{k}} \pi'_{0}(s)=\sum_{s\in R_{k}} \lambda(s)$ for all $k=1,\dots ,K$. Assume that the $k'$-th recurrent class has highest entropy rate value. Since MDP is communicating, there exists a policy $\Bar{\mu} \in \Pi^S_\M$ inducing an MC with only one recurrent class $R_{k'}$. Then $\nabla H(\M^{\Bar{\mu}}) \geq \nabla H(\M^{\mu})$. Moreover, the policy $\Bar{\mu}$ can also be constructed by some feasible solution $\Bar{\gamma}(s,a)$ of constraints~(\ref{opt1:con1})-(\ref{opt1:con5}).  

Let $\gamma^{\star}(s,a)$ be the optimal solution of~(\ref{opt1:obj})-(\ref{opt1:con5}) 
and $\hat{\mu}^{\star}$ be the policy constructed by $\eqref{cpolicy1}$ based on   $\gamma^{\star}(s,a)$. From the analysis in last paragraph, we can assume without loss of generality that the MC $\M^{\hat{\mu}^{\star}}$ is a unichain.
Therefore, $\nabla H(\M^{\hat{\mu}^{\star}}) = V^\star$ regardless of initial distribution. Since $V^\star  \geq \nabla H(\M)$ regardless of initial distribution,
 we have $\nabla H(\M^{\hat{\mu}^{\star}}) = \nabla H(\M)$. 
Also, by Lemma~\ref{prop:com_irr}, $\M^{\hat{\mu}^{\star}}$ is irreducible. Since $\sum_{a \in A(s)}\gamma(s,a)$ is equal to the limit distribution for state $s$, $\sum_{a \in A(s)}\gamma(s,a)>0$ for all $s \in S$, which means that  $Rc=S$. This further implies that $\hat{\mu}^\star=\mu^\star$, which completes the proof. 
\end{proof}

Finally, we proceed to prove Proposition~\ref{prop:transientoptimal}.
The idea is to first translate the entropy rate problem to a total expected reward problem and then prove that the optimal solution of program \eqref{opt2:obj}-\eqref{opt2:con4} maximizes the total expect reward.

\begin{proof}[\bf Proof of Proposition~\ref{prop:transientoptimal}]
Let $Q=T_{k} \cup L_{k+1}= \hat{R}_{k+1} \setminus \hat{R}_k$.
Consider $\tilde{\M}_{k+1}=\hat{\M}_{k+1,\hat{R}_k}$ such that $\hat{\M}_{k+1,\hat{R}_k}$ is defined in \eqref{eq:newMDP}.
Moreover, a reward function $\tilde{\mathcal{R}}$ is equipped with $\tilde{\M}_{k+1}$ satisfying $\tilde{\mathcal{R}}(s,a)=\texttt{val}(s)$ for $s \in \hat{R}_{k}$ and $\tilde{\mathcal{R}}(s,a)=0$ for $s \in Q \cup \{ b\}$.

For any policy $\mu \in \Pi^{S}_{\tilde{\M}_{k+1}}$, 
let $\mu' \in \Pi_{\hat{\M}_{k+1}}^S$ be the policy such that $\mu'(s,a)=\hat{\mu}_{k}(s,a)$ for $s \in \hat{R}_k$ and $\mu'(s,a)=\mu(s,a)$ for $s \in Q$. 
We denote by $\mathcal{H}^{\mu'} \in \mathbb{R}^{|Q|}$ the entropy rate vector such that $\mathcal{H}^{\mu'}(s)$ is the entropy rate of MC $\hat{\M}_{k+1}^{\mu'}$ when the initial state is $s$. Now we prove that if every state of $Q$ is transient in $\hat{\M}_{k+1}^{\mu'}$, then for $\epsilon>0$ such that $\texttt{val} = \texttt{val}_k+\epsilon$, we have 
\begin{equation} \label{eq:vandh}
    \mathcal{H}^{\mu'}(s)+\epsilon=v_{\tilde{\M}_{k+1}}^{\mu}(s),\forall s \in Q.
\end{equation}
The entropy rate vector can be expressed equivalently as 
\[
\mathcal{H}^{\mu'}=\mathbb{P}_Q^{\mu'} \mathcal{H}^{\mu'} + \mathbb{P}_{\hat{R}_k}^{\mu'}(\texttt{val}-\epsilon\mathbf{e}),
\]
where $\mathbb{P}_Q^{\mu'}$ and $\mathbb{P}_{\hat{R}_k}^{\mu'}$ are the transition matrices from $Q$ to $Q$ and $\hat{R}_k$ of MC $\hat{\M}_{k+1}^{\mu'}$, respectively. 
When states in $Q$ are transient, $I-\mathbb{P}_Q^{\mu'}$ is invertible and we get
\begin{equation}\label{eq:transitionentropyvalue}
    \mathcal{H}^{\mu'}=(I-\mathbb{P}_Q^{\mu'})^{-1} \mathbb{P}_{\hat{R}_k}^{\mu'}(\texttt{val}- \epsilon\mathbf{e}).
\end{equation}
Let $\mathbb{P}_Q^\mu$ and $\mathbb{P}_{\hat{R}_k}^{\mu}$ be transition matrices from $Q$ to $Q$ and $\hat{R}_k$ of MC $\tilde{\M}^{\mu}_{k+1}$, respectively. From \cite[(7.1.6)]{puterman},  we have
\[
v_{\tilde{\M}_{k+1}}^{\mu,Q}=\mathbb{P}_Q^\mu v_{\tilde{\M}_{k+1}}^{\mu,Q} + \mathbb{P}_{\hat{R}_k}^{\mu} \texttt{val}.
\]
By $\mathbb{P}_Q^\mu=\mathbb{P}_Q^{\mu'}$, $\mathbb{P}_{\hat{R}_k}^{\mu}=\mathbb{P}_{\hat{R}_k}^{\mu'}
 $ and \eqref{eq:otherpropmiddle}, we know that (\ref{eq:vandh}) holds. 

We now prove that the optimal solution $\gamma^{\star}(s,a)$ of LP (\ref{opt2:obj})-(\ref{opt2:con4}) can generate policy that maximizes total expected reward $\tilde{\mathcal{R}}$ w.r.t. $\tilde{\M}_{k+1}$ by equation \eqref{eq:trans-enf}. To this end, for $s \in \hat{R}_k \cup \{b \}$, we add variables $\gamma(s,\Bar{a})$ and constraints
\begin{align}
 \gamma(s,\Bar{a}) - \sum_{t \in Q} \lambda(t,s) &\leq 0,   s \in \hat{R}_k,  \label{eq:add1}\\
     \gamma(b,\Bar{a}) - \sum_{s \in \hat{R}_k} \gamma(s,\Bar{a})& \leq 0 \label{eq:add2}
\end{align}
to (\ref{opt2:con1})-(\ref{opt2:con4}) and the objective function is changed to
\begin{equation}
    \sum_{s \in \hat{R}_k} \texttt{val}(s) \gamma(s,\Bar{a}).\label{newobj}
\end{equation}
From (7.2.18) in~\cite{puterman}, we know that the total expected reward maximization problem of $\tilde{\M}_{k+1}$ with $\tilde{\mathcal{R}}$ is equivalent to the new linear program~(\ref{opt2:con1})-(\ref{opt2:con4}), (\ref{eq:add1}), (\ref{eq:add2}), (\ref{newobj}). 
Theorem 7.2.18 in~\cite{puterman} proves the existence of optimal basic feasible solution of new-LP and the policy constructed by \eqref{eq:trans-enf} is stationary policy that maximizes total expected reward $\tilde{\mathcal{R}}$ for $\tilde{\M}_{k+1}$. 

 For any optimal solution $\gamma^{\star}(s,a)$ of the new-LP, one can argue  by contradiction that $\gamma^{\star}(s,\Bar{a})=\sum_{t \in Q} \lambda^{\star}(t,s)$. Therefore, the optimal value of~(\ref{newobj}) is equal to
\[
\quad \sum_{s \in \hat{R}_k} \texttt{val}(s) \gamma^{\star}(s,\Bar{a})
= \sum_{s \in \hat{R}_k} \sum_{t \in Q} \texttt{val}(s) \lambda^{\star}(t,s).
\]
Therefore, the optimal values of LP (\ref{opt2:obj})-(\ref{opt2:con4}) and new-LP (\ref{opt2:con1})-(\ref{opt2:con4}), (\ref{eq:add1}), (\ref{eq:add2}), (\ref{newobj}) are equivalent. It means that optimal solution of LP (\ref{opt2:obj})-(\ref{opt2:con4}) can generate policy that maximizes total expected reward $\tilde{\mathcal{R}}$ w.r.t. $\tilde{\M}_{k+1}$ by Equations \eqref{eq:trans-enf}. Since $\texttt{val}(t)>0$ for any $t \in \hat{R}_k$, $s \in Q$ can get positive total expected reward only when $s$ is transient in the induce MC. Thus every state in $Q$ is transient in MC $\hat{\M}^{\tilde{\mu}_{k+1}}_{k+1}$. Then from Equation~(\ref{eq:vandh}) we know regardless of initial state,
\[
\nabla H(\hat{\M}^{\tilde{\mu}_{k+1}}_{k+1}) = \sup_{\mu \in \Pi^{T}_{\hat{\M}_{k+1}} } \nabla H(\hat{\M}^{\mu}_{k+1}).
\]
From~\eqref{eq:transitionentropyvalue}, we know in \eqref{eq:nextval}, we have $\texttt{v}_\text{trans}(s)=\nabla H(\hat{\M}^{\tilde{\mu}_{k+1}}_{k+1})$ when the initial state is $s$. This completes the proof.
\end{proof}

\bibliographystyle{plain}
\bibliography{main}

\end{document}